\documentclass[a4paper]{cas-sc}

\usepackage[authoryear,longnamesfirst]{natbib}

\usepackage{mathtools} % Φορτώνει και το amsmath, διορθώνει το rcases
\usepackage{array}
\usepackage{nicematrix}
\DeclareMathOperator*{\argmin}{argmin}
\DeclareMathOperator*{\argmax}{argmax}
\usepackage{fancyhdr}
\usepackage[dvipsnames]{xcolor}
\usepackage{xurl}
\usepackage{graphicx}
\usepackage{booktabs}
\usepackage{alltt}
\usepackage{multirow}
\usepackage{rotating}
\usepackage{csquotes}
\usepackage{setspace}
\usepackage{indentfirst}
\usepackage[english]{babel}
\usepackage{epstopdf}
\usepackage{float}
\usepackage{arydshln}
\usepackage{xr-hyper}
\def\tsc#1{\csdef{#1}{\textsc{\lowercase{#1}}\xspace}}
\tsc{WGM}
\tsc{QE}
\begin{document}
\let\WriteBookmarks\relax
%\def\floatpagepagefraction{1}
%\def\textpagefraction{.001}

% Short title
\shorttitle{Bayesian optimization of combination tumor therapies}    

% Short author
\shortauthors{Lampropoulos et al.}  

% Main title of the paper
\title [mode = title]{A Bayesian-optimization framework coupling a multiphase PDE tumor model to efficiently design combination therapy schedules}  

% Title footnote mark
% eg: \tnotemark[1]
\tnotemark[1] 

% Title footnote 1.
% eg: \tnotetext[1]{Title footnote text}
%\tnotetext[1]{Submitted to \emph{Computer Methods in Applied Mechanics and Engineering}.} 

% First author
%
% Options: Use if required
% eg: \author[1,3]{Author Name}[type=editor,
%       style=chinese,
%       auid=000,
%       bioid=1,
%       prefix=Sir,
%       orcid=0000-0000-0000-0000,
%       facebook=<facebook id>,
%       twitter=<twitter id>,
%       linkedin=<linkedin id>,
%       gplus=<gplus id>]

\author[1]{Ioannis Lampropoulos}[orcid=0009-0002-3092-8750]

% Corresponding author indication

% Footnote of the first author
%\fnmark[1]

% Email id of the first author
%\ead{}

% URL of the first author
%\ead[url]{}

% Credit authorship
% eg: \credit{Conceptualization of this study, Methodology, Software}
\credit{}

% Address/affiliation
\affiliation[1]{organization={School of Chemical Engineering, National Technical University of Athens},
            addressline={Zografou}, 
            city={Athens},
%          citysep={}, % Uncomment if no comma needed between city and postcode
            postcode={15772}, 
            country={Greece}}

\author[2]{Yorgos M. Psarellis}[orcid=0000-0003-2540-0505]\fnref{sanofi}

\author[1]{Michail Kavousanakis}[orcid=0000-0001-6621-9249]
\cormark[1]
% Footnote of the second author
%\fnmark[2]
% Email id of the second author
\ead{mihkavus@chemeng.ntua.gr}

% URL of the second author
%\ead[url]{}

% Credit authorship
%\credit{}

% % Address/affiliation
% \affiliation[2]{organization={},
%             addressline={}, 
%             city={},
% %          citysep={}, % Uncomment if no comma needed between city and postcode
%             postcode={}, 
%             state={},
%             country={}}

% % Corresponding author text
% \cortext[1]{Corresponding author}

% Address/affiliation
\affiliation[2]{organization={Department of Chemical and Biomolecular Engineering},
            addressline={Johns Hopkins University}, 
            city={Baltimore},
            postcode={}, 
            state={MD},
            country={USA}}

% Corresponding author text
\cortext[1]{Corresponding author}
\fntext[sanofi]{Currently at Sanofi US}

% Footnote text
%\fntext[1]{}

% For a title note without a number/mark
%\nonumnote{}

% Here goes the abstract
\begin{abstract}
Designing combination cancer therapies requires choosing not only which agents to combine but also their relative doses and timing -- decisions that critically shape the trade-off between efficacy and toxicity. High-fidelity mechanistic models of tumor growth, formulated as systems of coupled partial differential equations (PDEs), can in principle resolve how these scheduling choices interact with the tumor microenvironment, but each evaluation is computationally expensive, rendering brute-force exploration of the design space intractable. We present a Bayesian Optimization (BO) framework that treats a multiphase, vascularized, two-dimensional PDE tumor simulator as a black box and uses a Gaussian-process surrogate together with a Lower Confidence Bound acquisition policy to find schedules that maximize therapeutic outcomes within a small budget of expensive simulations. We orchestrate the COMSOL Multiphysics\textsuperscript{\textregistered} solver from Python using the MPh library and \texttt{GPyOpt}, producing a fully automated optimization loop in which a single simulation of $\sim$650 days of tumor evolution requires roughly 80 hours of wall time. The framework is applied to three clinically relevant scenarios: (i) a two-agent regimen (docetaxel + bevacizumab), (ii) a three-agent regimen (docetaxel + bevacizumab + external radiation) under reduced and full intensity, and (iii) a single-agent dose-fractionation problem in which efficacy is balanced against healthy-tissue toxicity through a weighted multi-objective formulation. The BO loop converges to clinically plausible optima with one to two orders of magnitude fewer simulations than an equivalent grid search, identifies docetaxel-induced radiosensitization as a decisive factor in the triple-therapy optimum, and recovers a fractionation regime consistent with clinical protocols when both efficacy and toxicity are considered. The framework is agnostic to the specifics of the underlying PDE model and provides a transferable methodology for design optimization of expensive engineered or biological simulators.
\end{abstract}

% Use if graphical abstract is present
%\begin{graphicalabstract}
%\includegraphics{}
%\end{graphicalabstract}

% Research highlights
\begin{highlights}
\item Automated COMSOL-Python framework for tumor therapy optimization.
\item Bayesian optimization cuts PDE evaluations by an order of magnitude.
\item Quantifies docetaxel-induced radiosensitization in triple therapy.
\item Identifies schedules balancing tumor control and healthy-tissue toxicity.
\end{highlights}

% Keywords
% Each keyword is seperated by \sep
\begin{keywords}
Bayesian optimization \sep Gaussian processes \sep PDE-constrained design \sep multiphase tumor model \sep combination chemotherapy \sep digital twin
\end{keywords}

\maketitle

\section{Introduction}\label{sec:Intro}

% \begin{itemize}
%     \item \textbf{Importance: cancer \& chemo}

Cancer remains one of the most challenging diseases to treat, in part because its biological mechanisms involve multiple interacting pathways, redundancy, and the rapid emergence of resistance~\cite{general_info1}. To counter this complexity, combination therapies that pair agents with distinct mechanisms of action have become a cornerstone of modern oncology, with documented efficacy across breast~\cite{genprot1}, lung~\cite{genprot2}, gynecological~\cite{genprot3} and head-and-neck cancers~\cite{genprot4}. A representative example, and the focus of the present study, is the combination of anti-VEGF (anti-angiogenic) agents with cytotoxic chemotherapy: bevacizumab paired with paclitaxel or docetaxel has been shown to improve outcomes relative to monotherapy in several cancers~\cite{sandler:2006,dhillon:2012,wright:2006,ishikura:2019} and is an established element of standard treatment protocols~\cite{chemprot1,chemprot2,chemprot3}.

The design space expands further when different therapeutic modalities, such as chemotherapy and radiotherapy, are combined~\cite{articleAnsari, Hainsworth_Spigel_Greco_Shipley_Peyton_Rubin_Stipanov_Meluch_2011, ec65835e01b44d72b24315a61643585d}. Such multimodal regimens are already employed clinically; for example, taxane-based chemotherapy combined with radiotherapy is part of established protocols for non-small-cell lung cancer~\cite{radioprot1,radioprot2}. Multimodality, however, introduces a new layer of design choices: in addition to selecting the agents themselves, one must specify their relative doses, sequence, and timing. These choices critically determine the balance between efficacy and toxicity.

Mechanistic models that resolve tumor dynamics in space and time -- in particular partial differential equation (PDE) models -- offer a principled way to explore how scheduling choices interact with the tumor microenvironment to shape outcomes such as tumor shrinkage and recurrence~\cite{mi2014prediction, liu2023mathematical, li2024mathematical}. PDE-based formulations are well suited to capture spatially structured processes such as angiogenesis~\cite{Jiang2020,Fan2012} and metastatic seeding~\cite{Lorusso2008, Neophytou2021}. The fidelity of these simulators, however, comes at a computational cost that places systematic optimization out of reach for brute-force approaches~\cite{Lampropoulos2023}. This bottleneck has limited the practical utility of high-fidelity in-silico models for treatment design, personalized digital twins, and other downstream translational applications~\cite{Harrer2019, Pappalardo2018, Viceconti2016, Moingeon2023, Kleeberger2025, SantaAnaTellez2023, Kaul2022, Cellina2023}.
To overcome this bottleneck, we couple a mechanistic PDE-based tumor simulator with a data-efficient probabilistic surrogate, and use Bayesian Optimization (BO) to navigate the treatment-design space. In our workflow the PDE model is treated as a black box: the BO loop iteratively proposes dosing schedules, evaluates them with the high-fidelity simulator, and uses the observed outcomes to update a Gaussian-process (GP) surrogate that guides further search. BO is particularly well suited to this setting because it is non-parametric, uncertainty-aware, and explicitly designed to minimize the number of expensive evaluations~\cite{Shahriari2016, Frazier2018, Rasmussen2005-ij}. It has a strong track record for the optimization of costly black-box systems in chemistry and biology~\cite{Psarellis2023, Psarellis2025, Manoj2025, Merzbacher2023, Shields2021, Thompson2024}, and an emerging body of work has begun to apply it to therapeutic dose and schedule design~\cite{Takahashi2020, Willard2024, Krishnamoorthy2022, Takahashi2021}. 
A further feature of our approach is its natural extension to multi-objective optimization. Combination therapies present competing objectives -- most prominently, maximizing efficacy while minimizing toxicity~\cite{Thall2004, Braun2010}. BO supports principled exploration of these trade-offs through weighted scalarization or Pareto-front construction, accounting for interactions between mechanisms of action rather than assuming independent dose-response curves~\cite{VanderHeiden2011, Chargari2016, Lee2007}.
The principal contributions of this work are as follows.
\begin{itemize}
    \item We integrate a spatially resolved, vascularized, multiphase PDE tumor model with an uncertainty-aware BO loop, enabling efficient treatment-schedule design when each PDE evaluation is computationally expensive (here, $\sim$80 h per simulation).
    \item We apply a bi-objective extension of the BO loop to expose efficacy--toxicity trade-offs for combinations of anti-angiogenic, cytotoxic, and radiotherapeutic modalities, producing interpretable, Pareto-optimal treatment regimes.
    \item We show that BO-guided exploration uncovers biologically meaningful trade-offs and regimen features -- such as the dominant role of docetaxel-induced radiosensitization -- that align with mechanistic insight and motivate testable hypotheses for experimental validation.
\end{itemize}

The remainder of the manuscript is organized as follows. Section~\ref{sec:moddev} describes the PDE-based tumor simulator that produces the high-fidelity evaluations. Section~\ref{sec:Methods} presents the Bayesian optimization methodology and the coupling between Python and the PDE solver. Section~\ref{sec:Results} reports three case studies (double, triple and single-agent regimens). Section~\ref{sec:Conclusions} discusses the implications and limitations of the framework. 

\section{Multiphase PDE tumor model}\label{sec:moddev}

Before introducing the optimization methodology, we summarize the mechanistic PDE-based tumor model that produces the high-fidelity (and computationally expensive) evaluations used by the optimizer. The model is a multiphase continuum formulation in two spatial dimensions and is described in full detail in previous work~\cite{Lampropoulos2023, lampropoulos2025modeling}; we restrict the present description to the elements required to interpret the results. A complete specification of the kinetic source terms is provided in the Supplementary Information (Sec.~I).

\begin{figure}[ht!]
   \centering
   \includegraphics[width=0.95\linewidth]{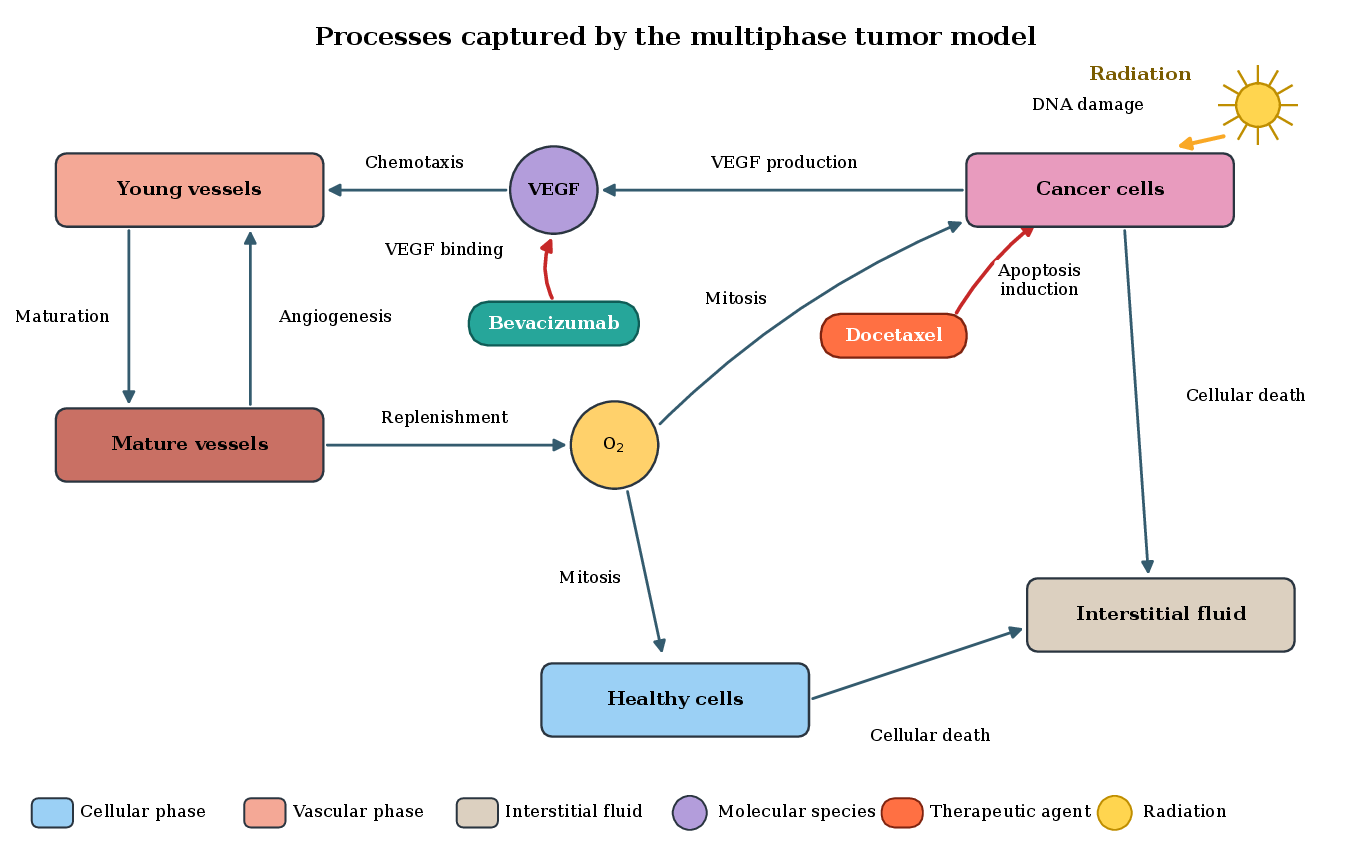}
   \caption{Schematic of the processes captured by the multiphase tumor
   model. Boxes denote the five volumetric phases (cancer, healthy,
   young and mature vessels, interstitial fluid); circles denote the
   molecular species (VEGF, O$_2$). Dark arrows mark the biological
   processes that couple the phases and species. The three therapeutic
   interventions considered in this work are overlaid: docetaxel
   (cytotoxic, induces apoptosis of cancer cells), bevacizumab
   (anti-angiogenic, binds and neutralizes VEGF) and external radiation
   (induces DNA damage in cancer cells).}
   \label{fig:model_schematic}
\end{figure}

The model represents five non-mixable volumetric phases, each with its own volume fraction, velocity field $\vec{u}_i=(u_i,v_i)$ and pressure $p_i$:
\begin{itemize}
    \item Healthy cells ($i=h$).
    \item Cancer cells ($i=c$).
    \item Young vessels ($i=yv$).
    \item Mature vessels ($i=mv$).
    \item Interstitial fluid ($i=int$).
\end{itemize}
Coupled to these, the model includes four molecular species that mediate biological processes and deliver the therapies: oxygen, vascular endothelial growth factor (VEGF), bevacizumab and docetaxel. The molecular species do not contribute to the volumetric composition. The governing system consists of mass and momentum balances for the phases, reaction--diffusion equations for the molecular species and algebraic equations of state. A graphical illustration of all processes captured by our model is depicted in Fig.~\ref{fig:model_schematic}.

\subsection{Mass balance equations for the cellular phases} \label{sec:massbalancecell}

Assuming macroscopically uniform tissue density~\cite{liu2022uniformity}, the mass balance for each phase $i\in\{h,c,yv,mv,int\}$ reads
\begin{equation}\label{eq:genmassbalance}
  \frac{\partial \theta_i}{\partial t} + \nabla \cdot \left( \vec{u}_i \theta_i \right) +  \chi_{J} \nabla \cdot \left( \theta_i \nabla J  \right) = q_i,
\end{equation}
where $\theta_i$ is the volume fraction of phase $i$. The first term on the left represents local accumulation, the second advection by the phase velocity, and the third directed motion (chemotaxis) driven by a chemokine $J$. The right-hand side $q_i$ collects all interphase mass exchange and kinetic source/sink terms (proliferation, death, vessel formation, drug uptake).

On the inflow portion of the boundary $\partial \Omega$ (where $\vec{u}_i \cdot \vec{n} < 0$), we prescribe Dirichlet values equal to the initial volume fractions:
\begin{equation}\label{eq:bctheta}
  \theta_i = \theta_{i}(t=0), \quad i\in\{h,c,yv,mv,int\}.
\end{equation}

\subsection{Momentum balance}\label{mom_bal}
Tissue-scale flows operate in the creeping-flow regime ($Re<1$); inertial terms are neglected and the momentum balance for each phase reduces to
\begin{equation}\label{eq:mombalgeneral}
  \nabla \cdot \left( \theta_i \boldsymbol{\sigma}_i \right) + \vec{F}_i = \vec{0}, \quad i\in\{h,c,yv,mv,int\}.
\end{equation}
Treating each volumetric phase as a Newtonian viscous fluid with dynamic viscosity $\mu_i$, the stress tensor is
\begin{equation}\label{eq:sigmatensor}
  \boldsymbol{\sigma}_i = -p_i \boldsymbol{I} + \mu_i \left( \nabla \vec{u}_i + (\nabla \vec{u}_i)^T \right) -\tfrac{2}{3} \mu_i \left( \nabla \cdot \vec{u}_i \right) \boldsymbol{I}.
\end{equation}
The interphase force $\vec{F}_i$ collects pressure-driven surface forces and inter-phase drag,
\begin{equation}\label{eq:forcegeneral}
  \vec{F}_i = p_i \boldsymbol{I} \nabla \theta_i + \sum_{j\neq i} d_{i,j} \theta_i \theta_j \left( \vec{u}_j - \vec{u}_i \right),
\end{equation}
with $d_{i,j}$ a drag coefficient. Boundary conditions are traction-free on the external boundary for the cellular and vessel phases,
\begin{equation}\label{eq:bcmom}
  \boldsymbol{\sigma}_i \cdot \vec{n} = \vec{0}, \quad i\in\{h,c,yv,mv\},
\end{equation}
together with a no-flow condition for the interstitial fluid,
\begin{equation}\label{eq:bcmom2}
  \vec{u}_{int} = \vec{0}.
\end{equation}
Summing the mass balances over all phases yields the continuity relation used to recover the pressure field,
\begin{equation}\label{eq:conservmom}
  \sum_{i} \left[\nabla \cdot \left( \theta_i \vec{u}_i \right) + \sum_{J} \chi_J \nabla \cdot \left( \theta_i \nabla J \right) \right] = 0.
\end{equation}
Once the interstitial pressure $p_{int}$ is recovered, algebraic equations of state close the remaining phase pressures~\cite{hubbard2013multiphase}.

\subsection{Molecular species transport}
The four molecular species (oxygen, VEGF, bevacizumab, docetaxel) are governed by reaction--diffusion equations,
\begin{equation}\label{eq:nutrmb}
  D_{J} \nabla^2 J + s_{J} = 0, \quad J\in\{c,g,a,w\},
\end{equation}
with diffusion coefficient $D_J$ and net source term $s_J$, and Neumann (zero-flux) boundary conditions on $\partial\Omega$:
\begin{equation}\label{eq:nutrientbc}
  \nabla J \cdot \vec{n} = 0.
\end{equation}
Explicit expressions for the production, consumption, binding and pharmacokinetic terms in $s_J$ are given in the Supplementary Information (Sec.~I.C).

\section{Bayesian Optimization framework}\label{sec:Methods}
Each evaluation of the PDE model in Sec.~\ref{sec:moddev} is computationally expensive: a single simulation of one combination-therapy schedule resolves $\sim\!1.9\times 10^{6}$ degrees of freedom for $650$ non-dimensional time units (one unit corresponds to one day) and requires approximately 80 hours of wall-clock time on a 12-core AMD Ryzen 9 3900X workstation. A naive grid search over a single scheduling parameter at coarse resolution would already consume thousands of CPU-hours; the cost grows combinatorially in higher dimensions. Bayesian Optimization (BO) is well suited to this regime because it explicitly aims to identify the global optimum of an expensive, possibly noisy black-box objective using as few evaluations as possible~\cite{Shahriari2016, Frazier2018}.

We treat the PDE simulator as a black box. The BO loop iteratively proposes candidate schedules, evaluates them with the simulator, updates a probabilistic surrogate over the design space, and selects the next evaluation by maximizing an acquisition function that balances exploration and exploitation. The generic problem is
\begin{equation}\label{eq:opt_problem}
    \vec{x}^* = \argmax_{\vec{x}\in\mathcal{X}} F(\vec{x}),
\end{equation}
with $\vec{x}$ the vector of scheduling decision variables, $F$ the objective computed by the simulator, and $\mathcal{X}$ a bounded search domain.

\begin{figure}[!ht]
    \centering
    \includegraphics[width=0.85\linewidth]{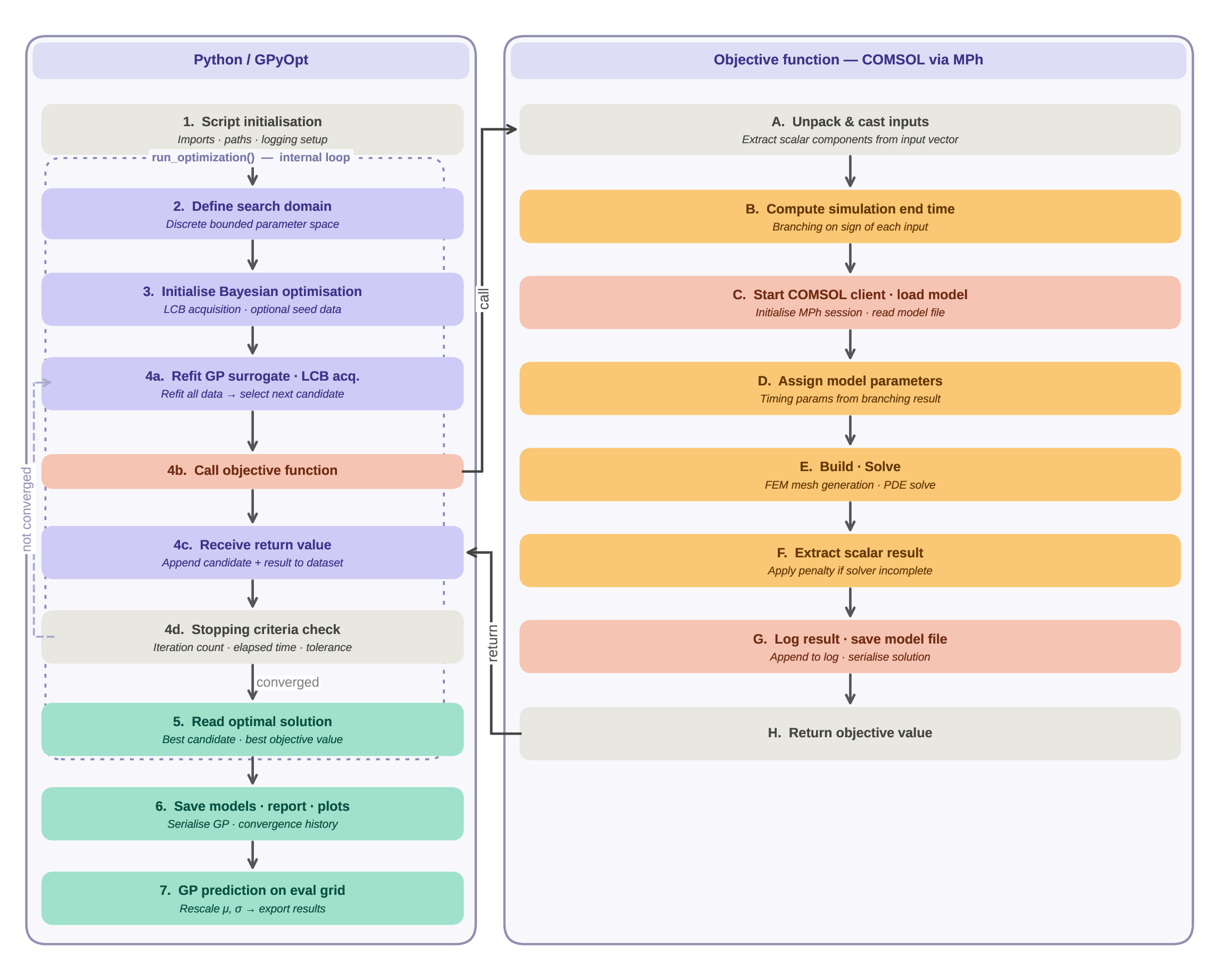}
    \caption{Optimization algorithm flow chart. Purple indicates the GPyOpt/BO logic, coral boxes represent the COMSOL/MPh interface, amber shading marks the calculations carried out within the objective function, and teal denotes the post-processing steps.
    }
    \label{fig:bo_pipeline}
\end{figure}

The optimization framework outlined above is encapsulated in the flowchart shown in Fig.~\ref{fig:bo_pipeline}. Its components are examined in greater detail in the following paragraphs.

\subsection{Surrogate and acquisition}

We adopt a Gaussian process (GP) surrogate with a squared-exponential (radial basis function, RBF) kernel for its smoothness and good empirical performance on tumor-growth landscapes. Length-scale hyperparameters are re-estimated by maximum marginal likelihood after each new observation. The acquisition function is the Lower Confidence Bound (LCB),
\begin{equation}\label{eq:LCB}
    \mathrm{LCB}(\vec{x})=\mu(\vec{x})-\kappa\,\sigma(\vec{x}),
\end{equation}
with $\mu(\vec{x})$ and $\sigma(\vec{x})$ the GP posterior mean and standard deviation, and $\kappa$ controlling the exploration--exploitation trade-off (small $\kappa$ favours exploitation, large $\kappa$ favours exploration). Throughout this work we use $\kappa=0.1$, an exploitation-leaning choice motivated by the high per-evaluation cost. To maximize $F$ in Eq.~\eqref{eq:opt_problem} using the LCB, we equivalently minimize $-F$.

BO proceeds in batches of size one: after each completed simulation, the new observation $(\vec{x}_{\mathrm{next}}, F(\vec{x}_{\mathrm{next}}))$ updates the surrogate, and the acquisition function is then maximized over $\mathcal{X}$ to propose the next $\vec{x}$. The loop terminates when either a pre-set evaluation budget is reached or the incumbent best ceases to improve. The best observed schedule is returned as the recommendation. BO is implemented through \texttt{GPyOpt}~\cite{gpyopt2016}; COMSOL Multiphysics\textsuperscript{\textregistered} (Finite Element Method) is orchestrated from Python through the \texttt{MPh} library~\cite{mphlib}. The implementation details are provided in the Supplementary Information (Sec.~V).

\subsection{Decision variables and objectives}

We consider three families of optimization problems.

\paragraph{Double therapy (docetaxel + bevacizumab).} The decision variable is the offset between the first administration of the two agents,
\begin{equation}\label{eq:deltat}
    \vec{x} = \Delta t = t_{\mathrm{inj},1}^{\mathrm{doc}} - t_{\mathrm{inj},1}^{\mathrm{bev}},
\end{equation}
with $\Delta t\in[-106,106]\,\mathrm{d}$.

\paragraph{Triple therapy (docetaxel + bevacizumab + external radiation).} The decision vector contains two relative offsets,
\begin{align}
    \Delta t_1 &= t_{\mathrm{inj},1}^{\mathrm{doc}} - t_{\mathrm{inj},1}^{\mathrm{bev}}, \label{eq:deltat1}\\
    \Delta t_2 &= t_{\mathrm{rad},1} - t_{\mathrm{inj},1}^{\mathrm{bev}}, \label{eq:deltat2}
\end{align}
with $(\Delta t_1,\Delta t_2)\in[-84,84]^2\,\mathrm{d}$.

For the combined regimens, the objective is the Surface Coverage Mitigation ($SCM$), measuring the relative reduction in cancer-cell coverage with respect to an untreated reference at time $t$:
\begin{equation}\label{eq:chapCOMBO_SCM}
    SCM(t) = \left(1 - \frac{\iint_S \theta_c^{\mathrm{tr}}(x,y,t)\,\mathrm{d}x\,\mathrm{d}y}{\iint_S \theta_c^{\mathrm{un}}(x,y,t)\,\mathrm{d}x\,\mathrm{d}y}\right)\cdot 100\%.
\end{equation}
$SCM$ is evaluated 360 days after the last therapeutic injection,
\begin{equation}\label{eq:opt_problem1}
    \vec{\Delta t}^{\,*} = \argmax_{\vec{\Delta t}} SCM(\vec{\Delta t}).
\end{equation}

\paragraph{Single therapy (docetaxel only).} We optimize the splitting of a fixed total dose $w_{\mathrm{total}}$ over $N_{\mathrm{doc}}$ sessions of equal dose $w_{\mathrm{inj}}=w_{\mathrm{total}}/N_{\mathrm{doc}}$,
\begin{equation}\label{eq:opt_problem2}
    N_{\mathrm{doc}}^{*} = \argmax_{N_{\mathrm{doc}}}\, \vec{F}(N_{\mathrm{doc}}).
\end{equation}
Two metrics enter the vector-valued objective $\vec{F}$. The first is the tumor surface coverage,
\begin{equation}\label{eq:surfaceintegral}
    SC = \left(\frac{1}{S}\iint_S \theta_{c}\,\mathrm{d}x\,\mathrm{d}y\right)\cdot 100\%,
\end{equation}
and the second is healthy-tissue recession, used as a proxy for toxicity,
\begin{equation}\label{eq:surfaceintegraltox}
    HTR = \left(1 - \frac{\iint_S \theta_{h}\,\mathrm{d}x\,\mathrm{d}y}{\iint_S \theta_{h}(t=0)\,\mathrm{d}x\,\mathrm{d}y}\right)\cdot 100\%.
\end{equation}
Bi-objective optimization combines the two through a weighted scalarization (Section~\ref{sec:MOBO}).

\subsection{Computational setup}\label{sec:compsetup}

The PDE system is solved on a two-dimensional circular domain of non-dimensional radius 30, meshed with an unstructured Delaunay triangulation. All simulations start from a single cancerous lesion embedded in healthy tissue. A typical combination-therapy simulation comprises $\sim\!1.9\times 10^{6}$ degrees of freedom and $\sim\!80\,\mathrm{h}$ of wall time on a 12-core AMD Ryzen 9 3900X workstation. The optimization loop runs entirely in Python; \texttt{GPyOpt}~\cite{gpy_sheffieldml} hosts the GP surrogate and acquisition optimizer, and the \texttt{MPh} library~\cite{mph_pypi} drives the COMSOL solver and retrieves results. Representative implementation snippets are provided in the Supplementary Information (Sec.~V).

\section{Results}\label{sec:Results}

\subsection{Double therapy: optimizing the relative start time}\label{sec:double}
Docetaxel is administered in six sessions every three weeks at $175\,\mathrm{mg/m^2}$ per session~\cite{chemprot1, chemprot2}, spanning $106$ days. Bevacizumab follows the same calendar at $15\,\mathrm{mg/kg}$ per session~\cite{chemprot1, chemprot2}, also spanning $106$ days. The decision variable is the relative offset between the first administrations of the two agents, $\Delta t = t_{\mathrm{inj},1}^{\mathrm{doc}}-t_{\mathrm{inj},1}^{\mathrm{bev}} \in [-106,106]\,\mathrm{d}$; $\Delta t=-106$ corresponds to a fully serial schedule with docetaxel first, $\Delta t=+106$ to the reverse order, and $\Delta t=0$ to concomitant administration.

\begin{figure}[!ht]
   \centering
   \includegraphics[width=0.65\linewidth]{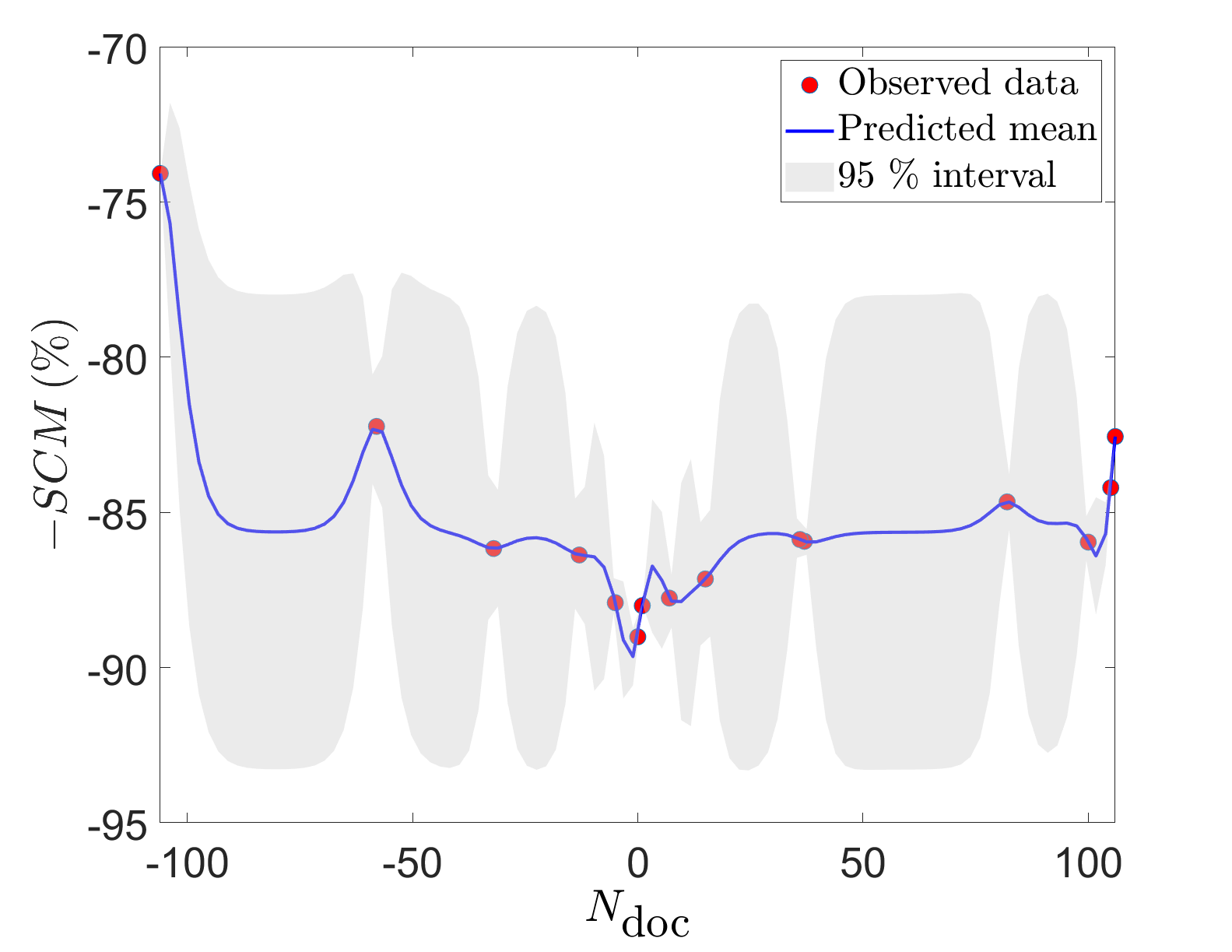}
   \caption{Gaussian-process surrogate learned by the BO loop for the docetaxel--bevacizumab schedule. The curve is the posterior mean of $-SCM$ as a function of $\Delta t$; red markers denote evaluated points (including the two boundary seeds); the shaded band is the posterior $\pm 1\sigma$ predictive uncertainty.}
   \label{fig:chapBO_1p_model}
\end{figure}

\begin{figure}[ht!]
   \begin{tabular}{cc}
   \includegraphics[width=0.5\linewidth]{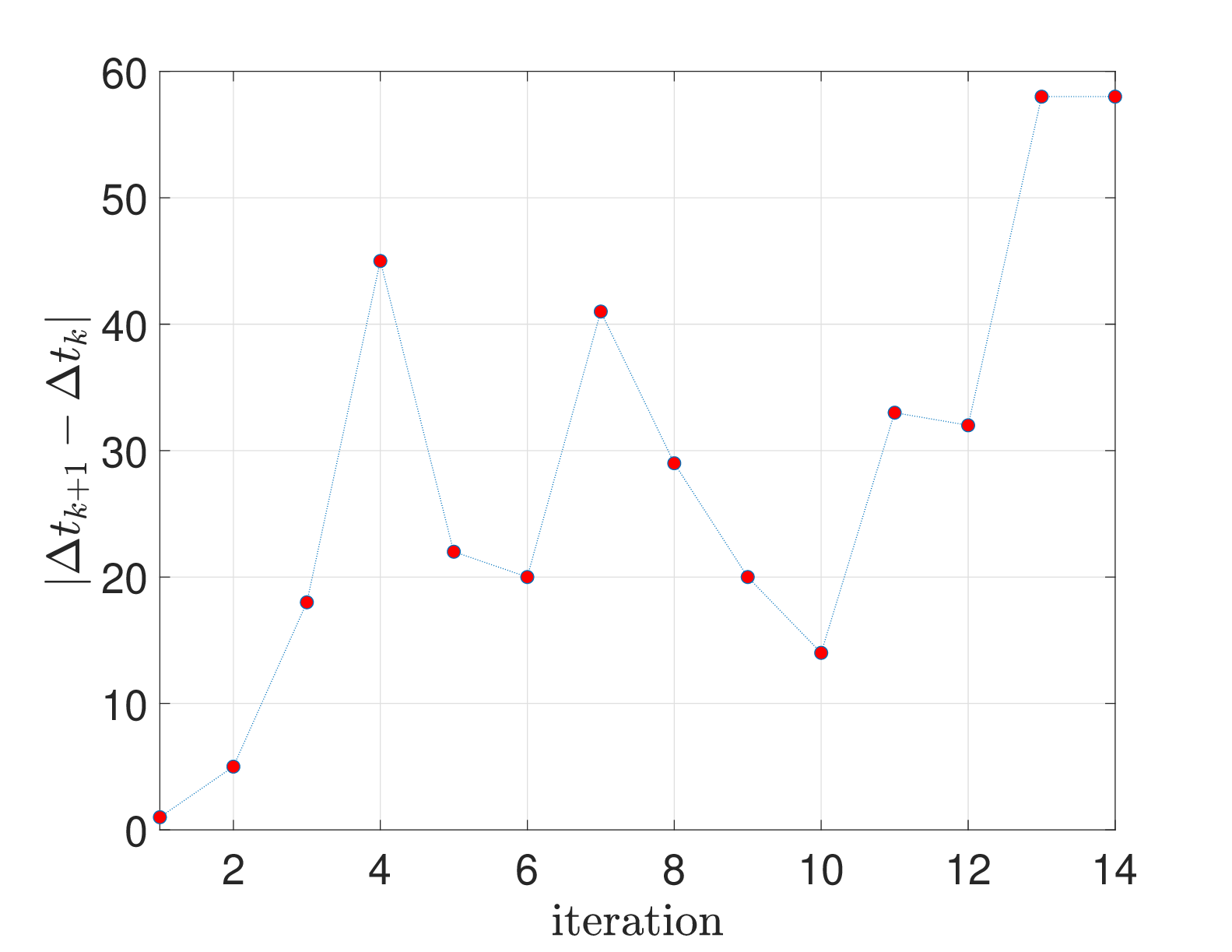}     &  \includegraphics[width=0.5\linewidth]{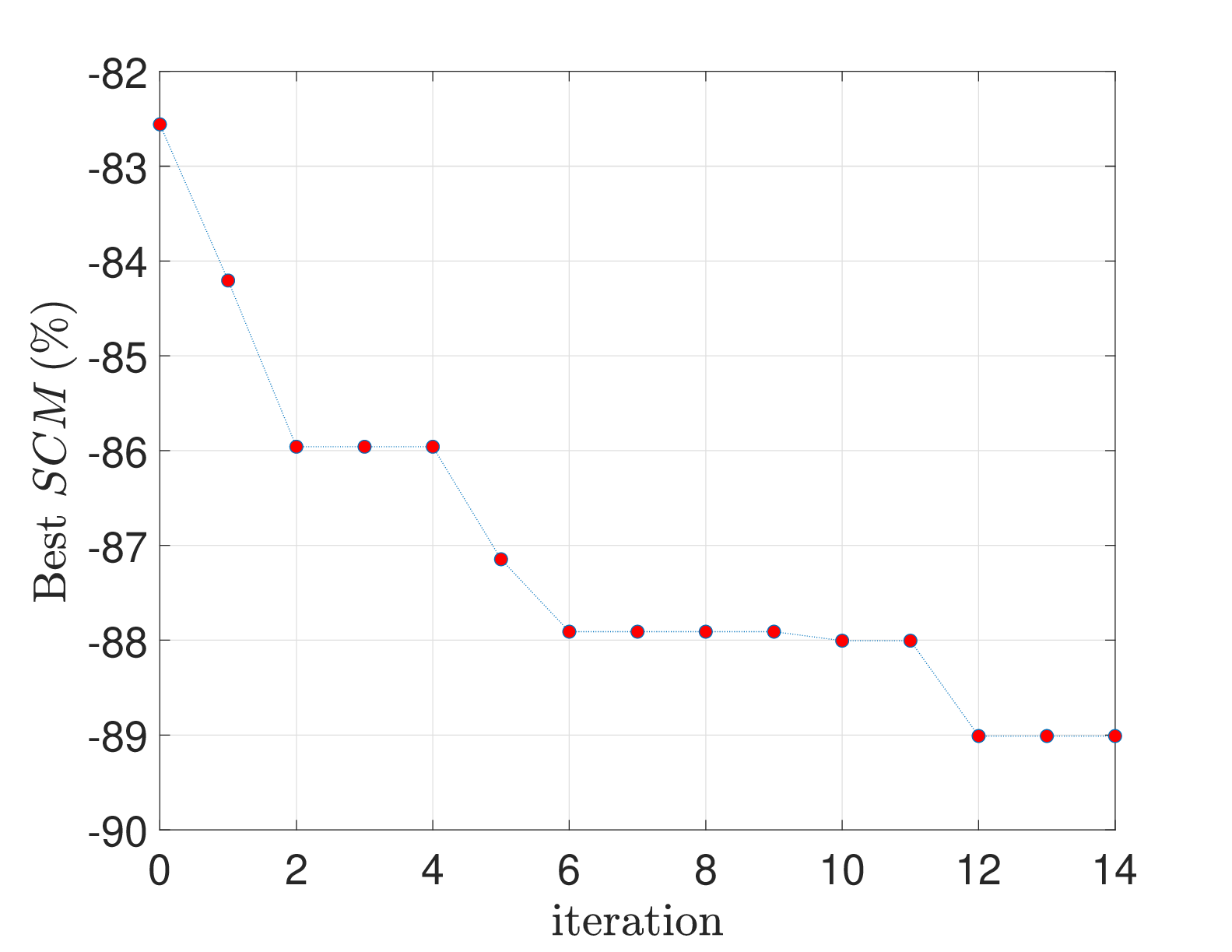} \\
   (a)     &  (b)
   \end{tabular}
   
   \caption{BO iteration diagnostics for the double-therapy case. \textbf{(a)} Distance between consecutive proposals, $|\Delta t_{k+1}-\Delta t_k|$, vs.~iteration. \textbf{(b)} Best observed $SCM$ vs.~iteration. The decreasing step size and rapidly saturating incumbent reflect the exploitation-focused acquisition.}
   \label{fig:double_iter}
\end{figure}

The optimization objective is $SCM$ evaluated 360 days after the final injection. With $t_{\mathrm{start}}=80$ and 21-day cycles, the sampling time is $t_{\mathrm{sample}}=545 + |\Delta t|$. The BO loop was seeded with the two boundary observations: $SCM=74.1\%$ at $\Delta t=-106$ and $SCM=82.6\%$ at $\Delta t=+106$. The acquisition function then drove the search toward the interior, refining around $\Delta t\approx 0$. The best observed schedule was $\Delta t\approx 0$ with $SCM\approx 89\%$ (Fig.~\ref{fig:chapBO_1p_model}), in agreement with the grid-based result of~\cite{Lampropoulos2023} but obtained at a small fraction of the computational cost. With the exploitation-leaning acquisition ($\kappa=0.1$), the proposal step size $|\Delta t_{k+1}-\Delta t_k|$ decreases monotonically as the algorithm focuses on the incumbent best (Fig.~\ref{fig:double_iter}). The result is consistent with clinical practice, where concomitant initiation of cytotoxic and anti-angiogenic agents is the recommended schedule for several indications.

\begin{figure}[ht!]
   \centering
   \includegraphics[width=0.95\linewidth]{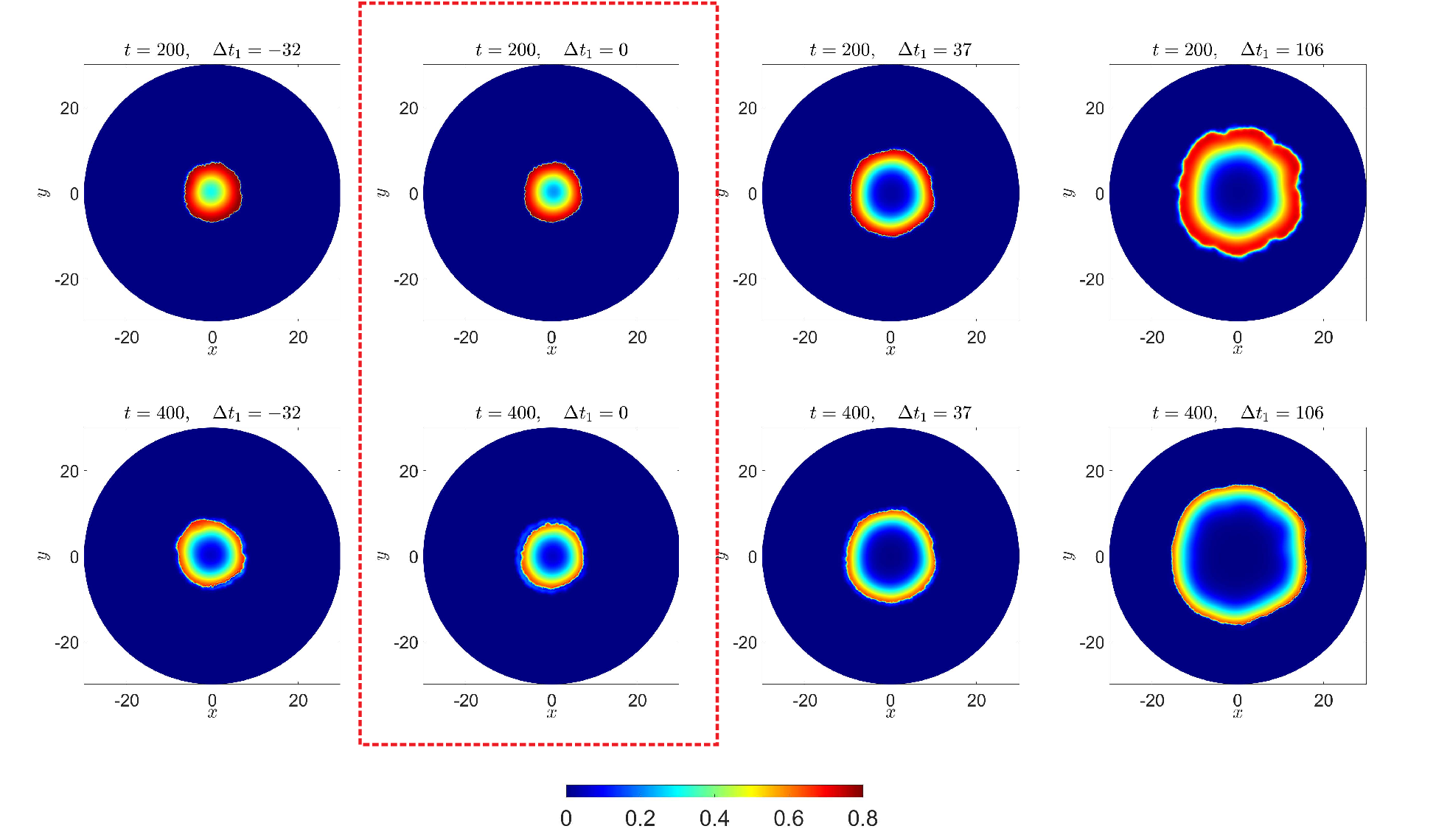}
   \caption{Cancer-cell volume fraction snapshots at $\Delta t\in\{-32,0,37,106\}$ and at $t=200$ (top row) and $t=400$ (bottom row) days. Near-concomitant schedules yield substantially smaller tumor coverage at long horizons than strongly offset schedules.}
   \label{fig:double_surf_eg}
\end{figure}

Spatial snapshots of the cancer-cell volume fraction at representative offsets $\Delta t\in\{-32,0,37,106\}$ (Fig.~\ref{fig:double_surf_eg}) illustrate the qualitative consequence of the schedule. Near-concomitant schedules suppress the tumor uniformly throughout the domain; strongly offset schedules yield larger total burden and rougher tumor boundaries. The mild non-monotonicity visible in the surrogate near $\Delta t\approx -30$ is consistent with discretization-level noise in the PDE simulator% (see Section~\ref{sec:limitations})
; the GP noise term absorbs this variance and produces well-calibrated uncertainty estimates in this region.

\subsection{Triple therapy: docetaxel + bevacizumab + external radiation}\label{sec:triple}
We next consider the three-agent regimen. The decision variables are the two relative offsets defined in Eqs.~\eqref{eq:deltat1}--\eqref{eq:deltat2}; $SCM$ remains the objective. Determining the sampling time $t_{\mathrm{sample}}$ is more involved than in the double-therapy case because both the sign and the relative magnitude of $\Delta t_1$ and $\Delta t_2$ enter the calendar. Five sign and ordering regimes arise:
\begin{enumerate}
    \item $\Delta t_1\geq 0$, $\Delta t_2\geq 0$,
    \item $\Delta t_1\geq 0$, $\Delta t_2 < 0$,
    \item $\Delta t_1 < 0$, $\Delta t_2\geq 0$,
    \item $\Delta t_1 < 0$, $\Delta t_2 < 0$ with $\Delta t_1\geq \Delta t_2$,
    \item $\Delta t_1 < 0$, $\Delta t_2 < 0$ with $\Delta t_1 < \Delta t_2$.
\end{enumerate}
The corresponding expressions for $t_{\mathrm{sample}}=t_{\mathrm{start}}+t_{\mathrm{elapsed}}+L$ are summarized in Table~\ref{tab:chBO_triple_sampling_time}.

\begin{table}[ht]
\centering
\caption{Expressions for the $SCM$-sampling time $L$ as a function of the sign and ordering of $(\Delta t_1, \Delta t_2)$.}
\begin{tabular}{ccc}
\toprule
$\mathrm{sgn}(\Delta t_1)$ & $\mathrm{sgn}(\Delta t_2)$ & $L$ \\
\midrule
$+$ & $+$ & $\max(\Delta t_1+42,\,\Delta t_2+18)$ \\
$+$ & $-$ & $\Delta t_1 + |\Delta t_2| + 42$ \\
$-$ & $+$ & $\max(|\Delta t_1|+42,\,|\Delta t_1|+\Delta t_2+18)$ \\
$-$ & $-,~\Delta t_1\geq\Delta t_2$ & $|\Delta t_1| + 42$ \\
$-$ & $-,~\Delta t_1<\Delta t_2$ & $|\Delta t_2| + 42$ \\
\bottomrule
\end{tabular}
\label{tab:chBO_triple_sampling_time}
\end{table}

We define the search box by enumerating all serial-ordering combinations of the three agents and taking the smallest box that contains them; the resulting bounds are $\Delta t_1, \Delta t_2\in[-84,84]\,\mathrm{d}$. The corresponding extreme schedules are listed in Table~\ref{tab:triple_bounds}.

\begin{table}[ht]
\centering
\caption{First-administration times $t_{\mathrm{inj},1}^{\mathrm{bev}}$, $t_{\mathrm{inj},1}^{\mathrm{doc}}$ and $t_{\mathrm{rad},1}$ (days) for the six serial orderings of the three agents, together with the corresponding $(\Delta t_1,\Delta t_2)$.}
\begin{tabular}{ccccc}
\toprule
$t_{\mathrm{inj},1}^{\mathrm{bev}}$ & $t_{\mathrm{inj},1}^{\mathrm{doc}}$ & $t_{\mathrm{rad},1}$ & $\Delta t_1$ & $\Delta t_2$ \\
\midrule
$80$  & $122$ & $164$ & $42$  & $84$  \\
$80$  & $140$ & $122$ & $60$  & $42$  \\
$122$ & $80$  & $164$ & $-42$ & $42$  \\
$140$ & $80$  & $122$ & $-60$ & $-18$ \\
$98$  & $140$ & $80$  & $42$  & $-18$ \\
$140$ & $98$  & $80$  & $-42$ & $-60$ \\
\bottomrule
\end{tabular}
\label{tab:triple_bounds}
\end{table}

When multiple therapeutic agents are present simultaneously in the tissue, their interactions must be modeled explicitly. Taxanes such as docetaxel are known to sensitize cells to radiation~\cite{golden2014taxanes}; in addition, tissue oxygen concentration modulates the response of cancer cells to radiation~\cite{oronsky2011six}. We therefore consider two variants of the triple-therapy model: a \emph{radiosensitization-positive} variant that retains both docetaxel- and oxygen-induced radiosensitization, and a \emph{radiosensitization-negative} variant that neglects the docetaxel contribution. Radiosensitization is modeled according to~\cite{lampropoulos2025modeling}; a detailed description is provided in the Supplementary Information (Sec.~I.D).%

\subsubsection{Reduced-intensity regimen}\label{sec:triple_reduced}

We first consider a reduced-intensity protocol with three docetaxel and bevacizumab sessions and 15 radiation fractions, designed to prevent ties at full elimination of the tumor (under the full protocol, several schedules drive $\theta_c\to 0$ within the simulation horizon and produce identical $SCM$, making them indistinguishable through Eq.~\eqref{eq:opt_problem1}). The full-intensity scenario, in which we replace $SCM$ by the time to elimination, is treated separately in Section~\ref{sec:triple_full}.

\begin{figure}[ht!]
   \centering
   \includegraphics[width=0.65\linewidth]{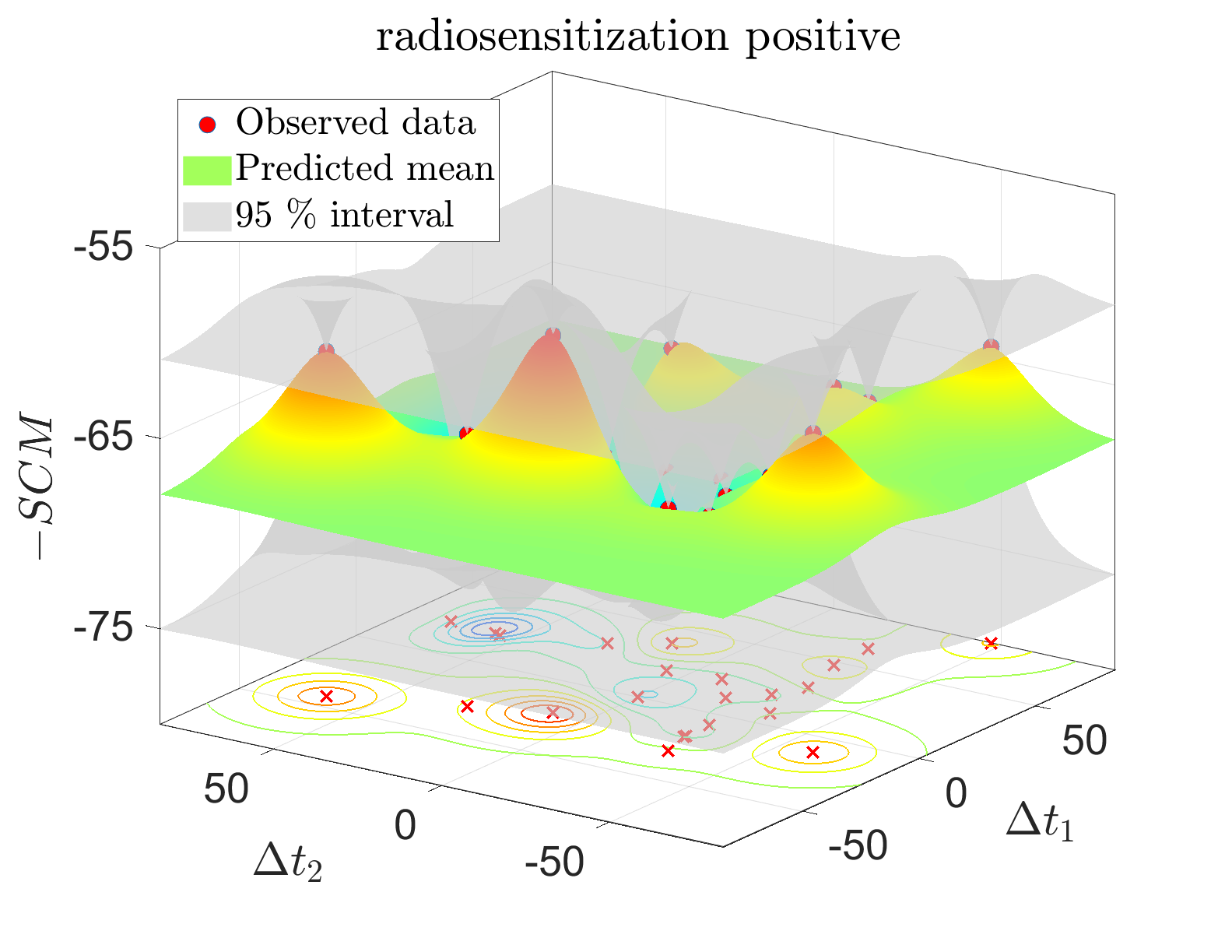}
   \caption{Posterior mean of $-SCM$ over $(\Delta t_1,\Delta t_2)$ for the radiosensitization-positive triple-therapy variant. Red markers denote BO iterates. A primary optimum lies near $(\Delta t_1,\Delta t_2)\approx(21,57)$ and a secondary, broader region of high efficacy spans $\Delta t_1\in[-20,40]$, $\Delta t_2\in[-40,10]$.}
   \label{fig:chap_BO_2p_model_SENS}
\end{figure}

\textbf{Radiosensitization-positive run.} The BO loop was initialized with no prior observations. After three iterations the loop converged to a local optimum at $(\vec{\Delta t}_{\mathrm{opt}}, SCM_{\mathrm{opt}})=((-18,-27),\,69.12\%)$. To rule out a local maximum, the loop was restarted with five additional seed points, after which the global incumbent moved to $(\vec{\Delta t}_{\mathrm{opt}}, SCM_{\mathrm{opt}})=((21,57),\,75.52\%)$. The corresponding optimization script is provided in the Supplementary Information (Sec.~V.B); we report only the summary statistics here. The corresponding GP posterior is shown in Fig.~\ref{fig:chap_BO_2p_model_SENS}.

Two regions of high efficacy emerge. The first, around $(\Delta t_1,\Delta t_2)\approx(21,57)$, corresponds to a schedule in which radiation is administered last; here the tumor has already been suppressed by chemotherapy, and docetaxel-induced radiosensitization further amplifies the effect of radiation on the residual tumor mass. The second is a broader region around $\Delta t_1\in[-20,40]$ and $\Delta t_2\in[-40,10]$, where the three agents are applied closer in time, with radiation either preceding or co-administered with chemotherapy. In this region the best observed schedule is $(\Delta t_1,\Delta t_2)=(12,-18)$, in which radiation hits the tumor early and chemotherapy maintains long-term suppression.

\begin{figure}[ht!]
   \centering
   \includegraphics[width=0.65\linewidth]{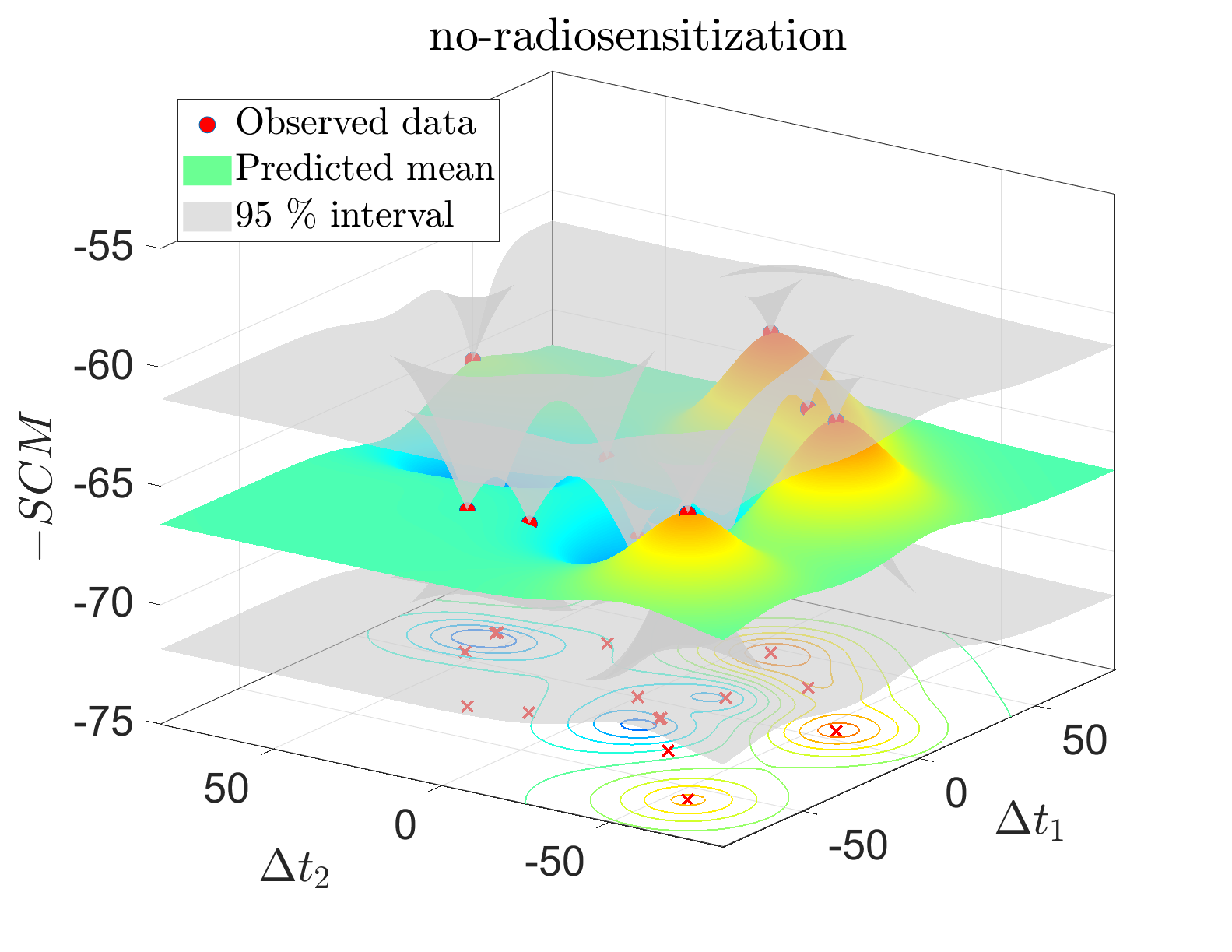}
   \caption{Posterior mean of $-SCM$ over $(\Delta t_1,\Delta t_2)$ for the radiosensitization-negative triple-therapy variant. The optimum is identified at $(\Delta t_1,\Delta t_2)\approx(21,57)$, the same scheduling regime as in Fig.~\ref{fig:chap_BO_2p_model_SENS}, but with a lower attained $SCM$.}
   \label{fig:chap_BO_2p_model_NOSENS}
\end{figure}

\textbf{Radiosensitization-negative run.} A second optimization, in which only oxygen-induced radiosensitization is retained, converges after five seedless iterations and a subsequent restart to $(\vec{\Delta t}_{\mathrm{opt}}, SCM_{\mathrm{opt}})=((21,57),\,69.83\%)$ (Fig.~\ref{fig:chap_BO_2p_model_NOSENS}). The corresponding setup follows the same implementation pattern reported in the Supplementary Information (Sec.~V.B).

Comparing the two GP posteriors, the topology of the surrogate is preserved between the two variants -- the same two regions are favored -- but the attained $SCM$ at the global optimum drops from $75.5\%$ to $69.8\%$ when docetaxel-induced radiosensitization is suppressed. We interpret this gap as a quantitative measure of the contribution of docetaxel-induced radiosensitization to long-term tumor control in the present scenario.

\subsubsection{Full-intensity regimen: time to elimination}\label{sec:triple_full}

\begin{figure}[ht!]
   \centering
   \includegraphics[width=0.65\linewidth]{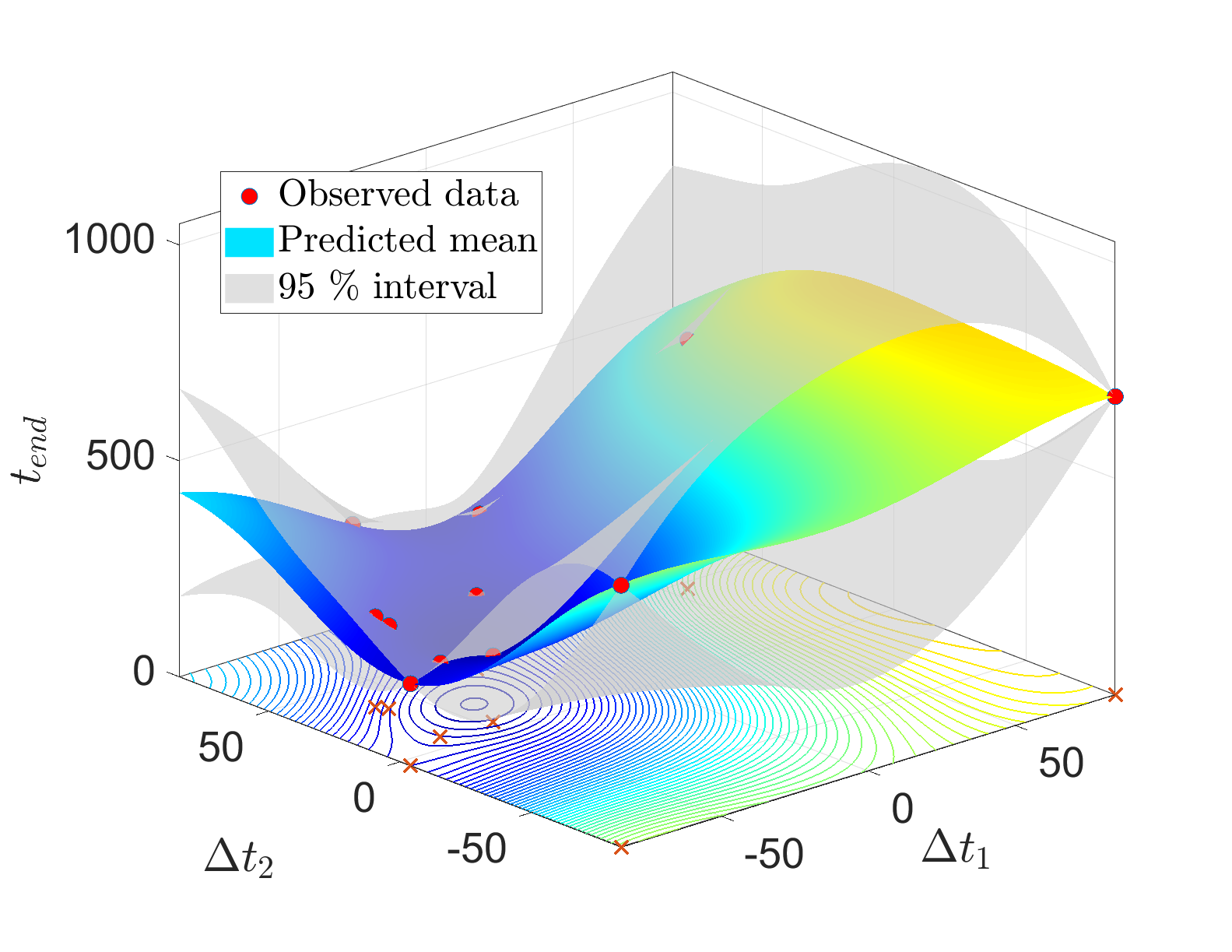}
   \caption{Posterior mean of $t_{\mathrm{end}}$ over $(\Delta t_1,\Delta t_2)$ for the full triple-therapy protocol with elimination-time objective. The sharp transition along the $(-84,-84)$--$(84,84)$ diagonal corresponds to the penalty applied to schedules that fail to eliminate the tumor.}
   \label{fig:chap_BO_2p_ELIM}
\end{figure}

Under the full protocol, several schedules drive the tumor to extinction within the simulation horizon. We therefore replace $SCM$ by the elimination time $t_{\mathrm{end}}$, defined as the first instant at which
\begin{equation*}
\iint_S \theta_c\,\mathrm{d}x\,\mathrm{d}y < 10^{-10}.
\end{equation*}
The objective becomes
\begin{equation}\label{eq:opt_problem_2}
    \vec{\Delta t}^{\,*} = \argmin_{\vec{\Delta t}} t_{\mathrm{end}}(\vec{\Delta t}).
\end{equation}
To prevent schedules that fail to eliminate the tumor from contaminating the surrogate, an additive penalty is applied:
\begin{equation*}
    t_{\mathrm{end}} \leftarrow t_{\mathrm{end}} + 300 \quad \text{if the elimination criterion is not met.}
\end{equation*}
The optimization identifies the optimum
$(\vec{\Delta t}_{\mathrm{opt}}, t_{\mathrm{end}}^{\mathrm{opt}})=((-47,6),\,153.83\,\mathrm{d})$, consistent with the chemotherapy-first / radiation-second pattern observed in the reduced-intensity case. The separation between docetaxel and radiation is larger here, an expected consequence of the higher cumulative docetaxel concentration achieved under the full protocol. The negative $\Delta t_2$ indicates that radiation precedes bevacizumab; this allows radiation to act before anti-VEGF-driven vascular pruning limits oxygen delivery, exploiting oxygen-induced radiosensitization. The GP posterior (Fig.~\ref{fig:chap_BO_2p_ELIM}) reveals a high-efficacy zone bounded by $\Delta t_1\lesssim 20$ and $\Delta t_2\gtrsim -20$. The sharp contrast between this zone and its neighborhood reflects the binary nature of the penalized objective: schedules below the $(-84,-84)$--$(84,84)$ diagonal generally fail to eliminate the tumor and incur the penalty.

\begin{figure}[ht!]
   \begin{tabular}{cc}
   \includegraphics[width=0.5\linewidth]{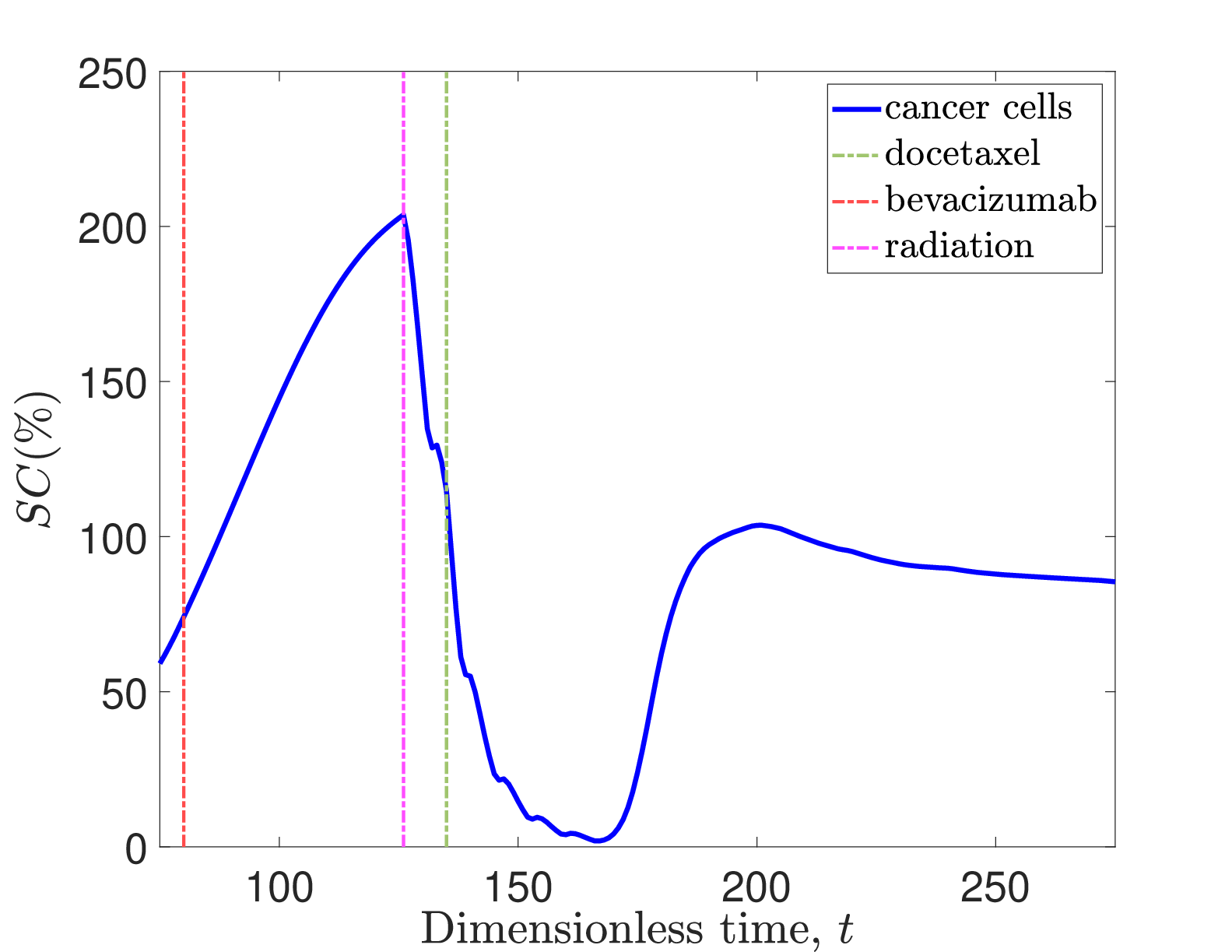}     & \includegraphics[width=0.5\linewidth]{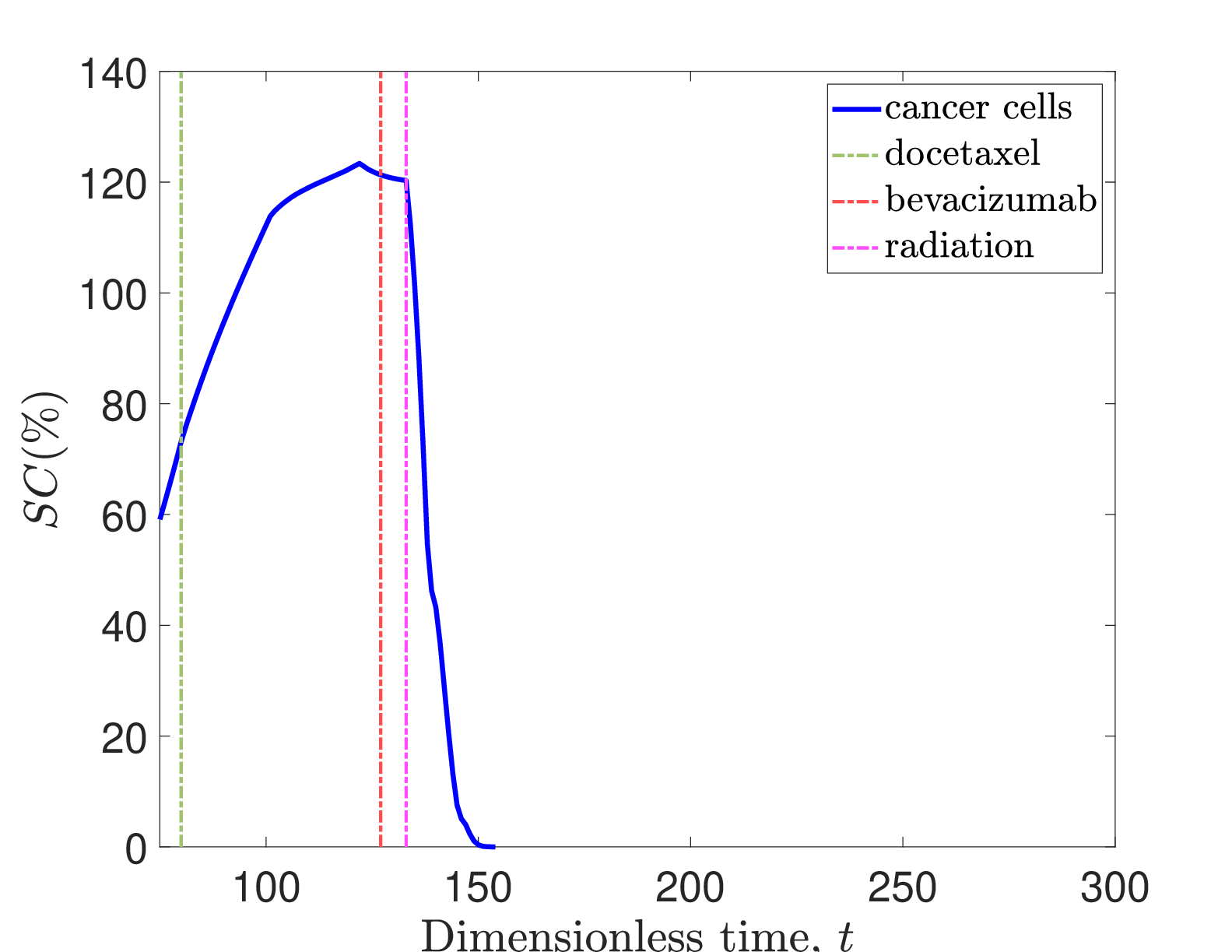} \\
   (a)     &  (b)
   \end{tabular}
   
   \caption{Cancer-cell surface coverage (\%) for the optimal schedule $(\Delta t_1,\Delta t_2)=(-47,6)$ (right) and the sub-optimal schedule $(55,46)$ (left). Color bars: bevacizumab (red), docetaxel (green), radiation (purple). Pre-treatment with docetaxel boosts the effect of radiation and accelerates tumor extinction; when bevacizumab leads, the residual cancer cells are contained but not fully eliminated.}
   \label{fig:triple1}
\end{figure}

The spatial dynamics underlying the optimum are illustrated in Fig.~\ref{fig:triple1}: pre-treatment with docetaxel sensitizes the tumor, and the subsequent radiation fraction eliminates the residual cancer cells. In contrast, when bevacizumab precedes the other agents, radiation does not fully eliminate the lesion and docetaxel only contains it.

\subsection{Single therapy: dose fractionation and the efficacy--toxicity trade-off}\label{sec:single}

We finally consider a single-agent regimen in which the decision variable is the number of docetaxel sessions $N_{\mathrm{doc}}$ at constant total dose $w_{\mathrm{total}}$, so that
\begin{equation*}
   w_{\mathrm{inj}} = \frac{w_{\mathrm{total}}}{N_{\mathrm{doc}}}.
\end{equation*}
This single-agent setting allows us to dissect the role of fractionation independently of inter-agent interactions, and to introduce a toxicity-aware bi-objective formulation.

\subsubsection{Efficacy-only optimization}\label{sec:single_efficacy}

\begin{figure}[ht!]
    \centering
    \includegraphics[width=0.65\linewidth]{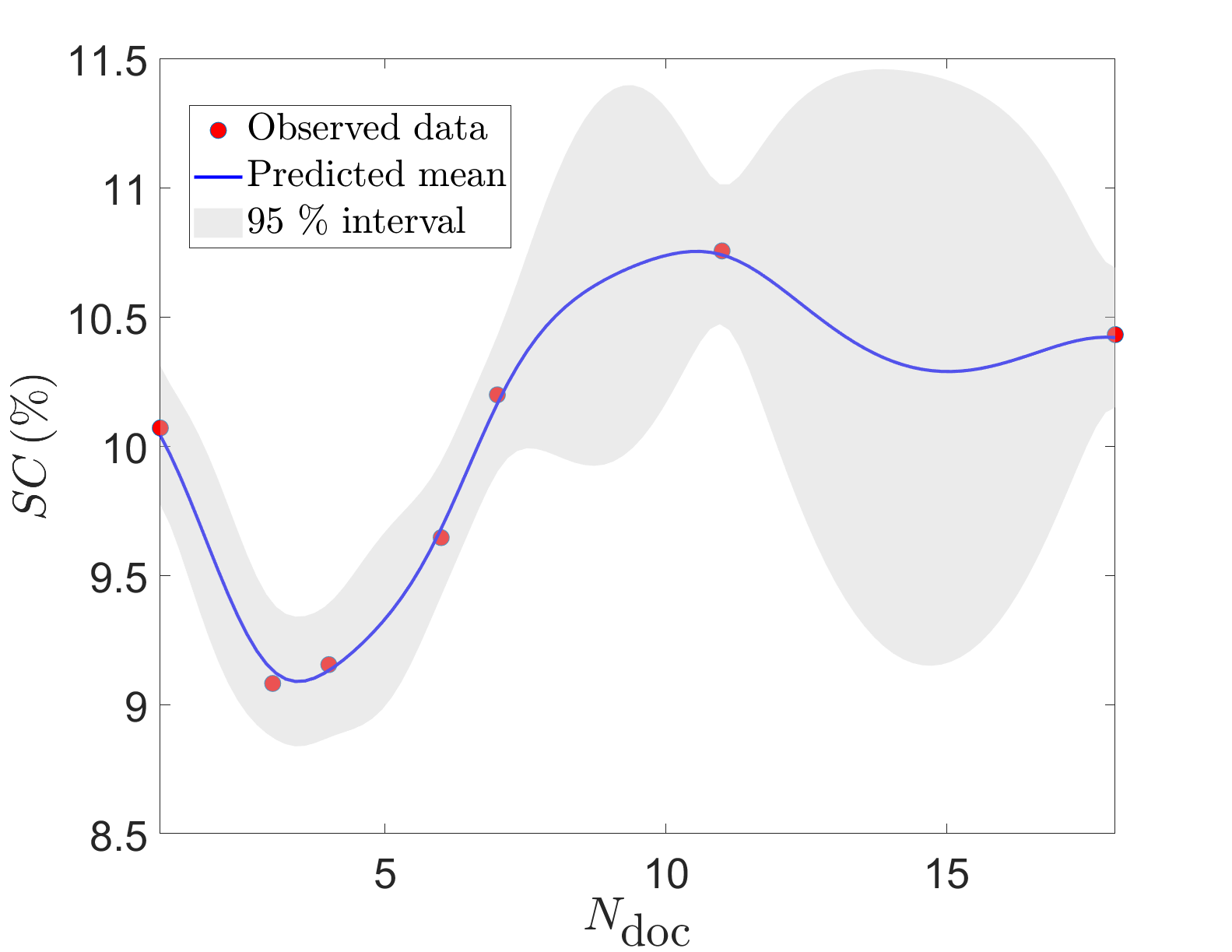}
    \caption{GP surrogate for the single-therapy efficacy-only problem. Posterior mean of $SC$ versus $N_{\mathrm{doc}}$; markers denote BO iterates. The optimum is identified at $N_{\mathrm{doc}}=3$.}
    \label{fig:SINGLE1}
\end{figure}

In the first sub-experiment, $SC$ is the sole objective. The BO loop was run without seed points; the implementation is summarized in the Supplementary Information (Sec.~V.E) and evaluates $N_{\mathrm{doc}}\in\{1,3,4,6,7,11,18\}$. The GP surrogate (Fig.~\ref{fig:SINGLE1}) identifies the optimum at $(N_{\mathrm{doc}},SC)=(3,\,9.08\%)$, with the neighbouring $N_{\mathrm{doc}}=4$ nearly as good. Efficacy degrades monotonically beyond $N_{\mathrm{doc}}\approx 6$, consistent with the intuition that excessive fractionation dilutes each session below the threshold at which it can outpace tumor regrowth. Standard clinical protocols~\cite{chemprot1, chemprot2} sit close to this optimum.

\begin{figure}[ht!]
    \centering
    \includegraphics[width=0.65\linewidth]{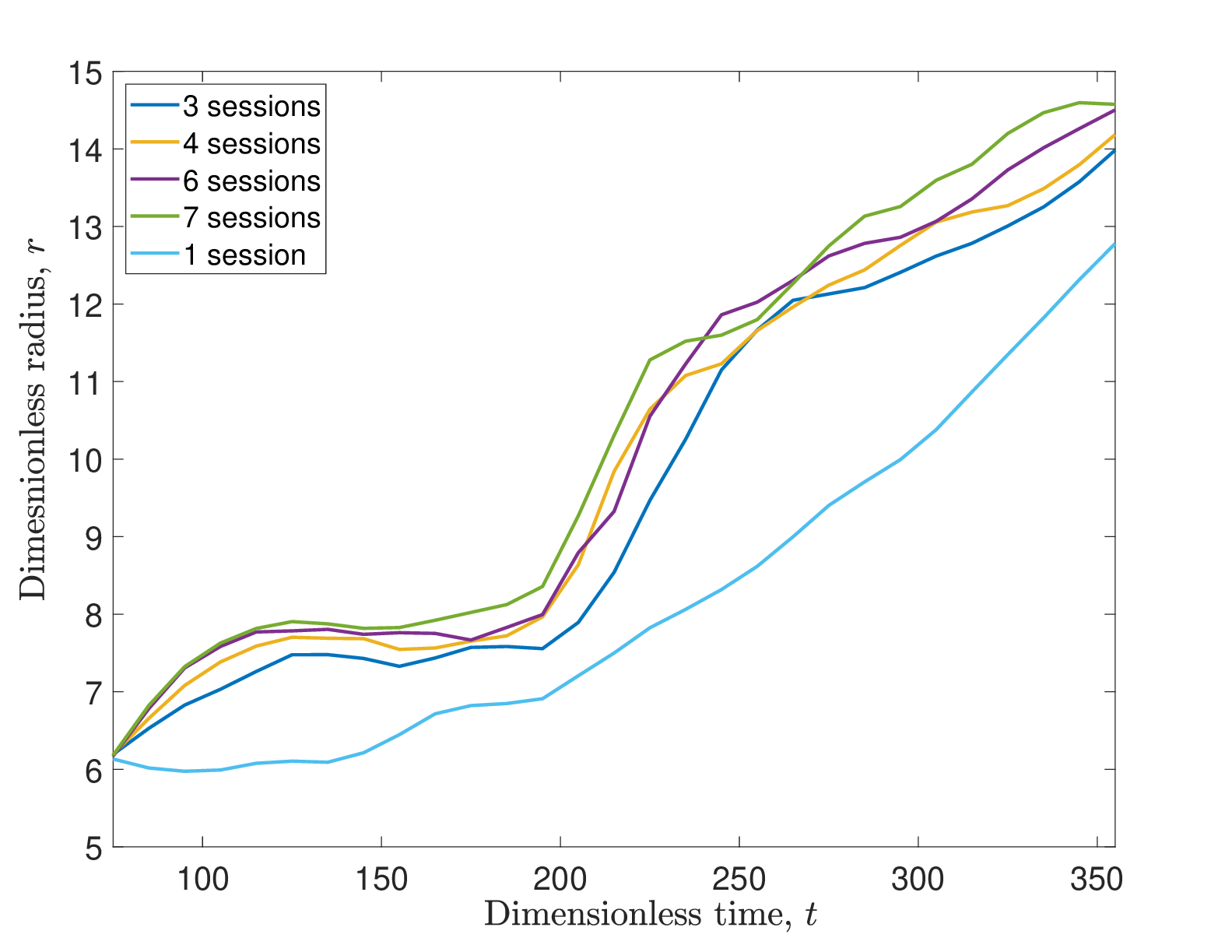}
    \caption{Dimensionless tumor radius $r$ (defined as the largest distance from the domain center at which $\theta_c=0.01$) for $N_{\mathrm{doc}}\in\{1,3,4,6,7\}$. Although $N_{\mathrm{doc}}=3$ minimizes the cancer-cell volume fraction, $N_{\mathrm{doc}}=1$ minimizes the tumor radius -- the cells are fewer in count but more diffusely distributed at $N_{\mathrm{doc}}=3$.}
    \label{fig:SINGLE1_1}
\end{figure}

A complementary observable, the tumor radius, is reported in Fig.~\ref{fig:SINGLE1_1}. The optimum $N_{\mathrm{doc}}=3$ minimizes the total cancer-cell content over the simulation horizon, but $N_{\mathrm{doc}}=1$ minimizes the radius -- consistent with a regimen that aggressively suppresses outward expansion but produces a higher residual cell density. 
This distinction motivates the bi-objective formulation that follows.

\begin{figure}[ht!]
    \centering
    \includegraphics[width=0.9\linewidth]{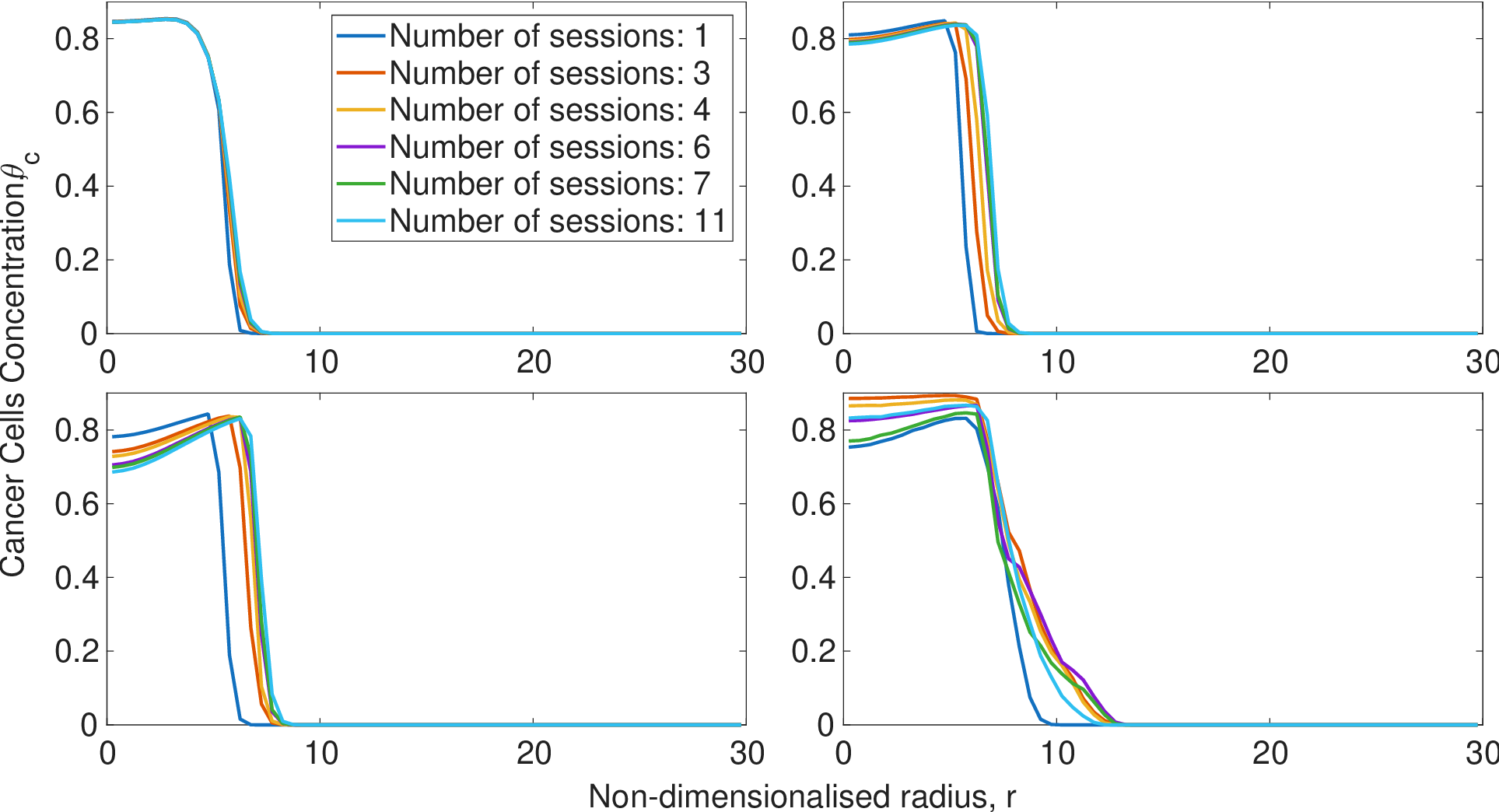}
    \caption{Cancer cell concentration, $\theta_c$, averaged over the tumor's non-dimensionalised radius, $r$, for $t=85$ (top left), $t=105$ (top right), $t=125$ (bottom left) and, $t=250$ (bottom right).}
    \label{fig:SINGLE1_2}
\end{figure}

The above is further illustrated in Fig.~\ref{fig:SINGLE1_2}, which presents the spatial profile of the mean cancer cell concentration for $r \in [0,30]$. Time point $t = 85$ was selected to characterize cancer cell behaviour immediately following the initial irradiation event. The subsequent time points, $t = 105$ and $t = 125$, provide insight into tumour dynamics as additional irradiation fractions are administered, whereas $t = 250$ depicts the radial distributions long after the completion of the therapeutic protocol.
A lesser degree of dose fractionation in radiotherapy appears to be associated with reduced tumor radius based on the simulations performed. Conversely, higher per-fraction radiation doses are generally associated with decreased tolerability in patients.
As mentioned above, these observations motivate the bi-objective approach described in the next paragraph.

\subsubsection{Bi-objective formulation: efficacy versus toxicity}\label{sec:MOBO}

To account for chemotherapy-induced toxicity to healthy tissue, an additional kinetic term is introduced in the healthy-phase source:
\begin{equation}
    q_h^{\mathrm{cyto}} = q_h - \phi_{\mathrm{tox}}\, k_{\mathrm{ap},c}\, \theta_h\, \frac{w}{w_m+w}\,\mathcal{H}\!\left(\theta_{\mathrm{int}}\frac{c}{c_p+c} - l_{\mathrm{cr},h},\, h_d\right),
\end{equation}
with $k_{\mathrm{ap},c}$ the docetaxel-induced apoptosis rate for cancer cells, $\phi_{\mathrm{tox}}$ a dimensionless coefficient capturing the increased resilience of healthy cells (which proliferate more slowly than cancer cells and have stronger DNA-repair machinery~\cite{mitchison2012proliferation, branham2004dna, xu2025second}), $w$ the local docetaxel concentration, $\mathcal{H}(m,n)=\tfrac{1}{2}[1+\tanh(m/n)]$ a smooth Heaviside approximation, $c_p$ the half-maximum oxygen concentration for mitosis, $l_{\mathrm{cr},h}$ the critical proliferation rate above which a cell is treated as highly proliferative, and $h_d$ the Heaviside smoothing parameter. With this term, healthy cells are assumed to die from docetaxel at a rate $k_{\mathrm{ap},h}=\phi_{\mathrm{tox}}\,k_{\mathrm{ap},c}$.

We combine $SC$ and $HTR$ through a weighted scalarization,
\begin{equation}\label{eq:scalarized}
    F(N_{\mathrm{doc}};w) = w\cdot SC(N_{\mathrm{doc}}) + (1-w)\cdot HTR(N_{\mathrm{doc}}),
\end{equation}
and run a separate BO loop for each of five weights $w\in\{0,\,0.3,\,0.5,\,0.7,\,1\}$. The optima are summarized in Table~\ref{tab:singleMOBO1}.

\begin{table}[ht]
    \centering
    \caption{Optimal fractionation $N_{\mathrm{doc}}^{*}$ and corresponding scalarized objective $F^{*}$ for five values of the weight $w$ in Eq.~\eqref{eq:scalarized}.}
    \begin{tabular}{cccccc}
    \toprule
    $w$ & $0$ & $0.3$ & $0.5$ & $0.7$ & $1$ \\
    \midrule
    $N_{\mathrm{doc}}^{*}$ & $14$ & $6$ & $6$ & $6$ & $3$ \\
    $F^{*}$ (\%)           & $28.55$ & $22.96$ & $19.16$ & $15.35$ & $8.89$ \\
    \bottomrule
    \end{tabular}
    \label{tab:singleMOBO1}
\end{table}

\begin{figure}[ht!]
    \centering
    \includegraphics[width=0.65\linewidth]{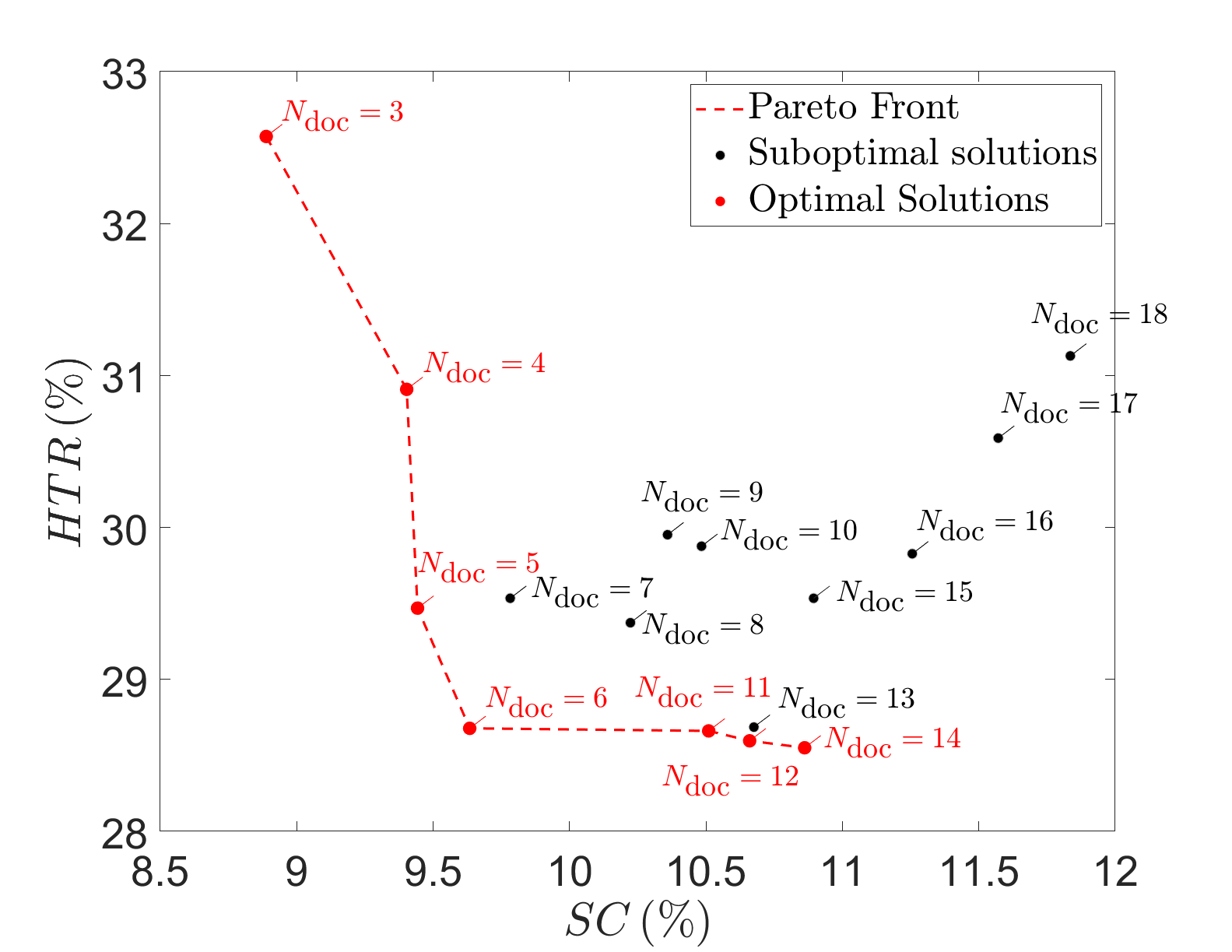}
    \caption{Pareto front for the single-therapy bi-objective problem. Points belonging to the Pareto-optimal set are highlighted in red; the dashed line traces the empirical front. Fractionation regimes $N_{\mathrm{doc}}\in[3,6]$ are optimal for tumor containment ($SC$); $N_{\mathrm{doc}}\in[11,14]$ are optimal for toxicity minimization ($HTR$). The clinically prescribed $N_{\mathrm{doc}}=6$ lies on the front.}
    \label{fig:pareto}
\end{figure}

At $w=0$ (toxicity-only), the optimizer settles on $N_{\mathrm{doc}}^{*}=14$, indicating that micro-dosing is preferable from the toxicity standpoint -- but only up to a point, beyond which further fractionation dilutes each session enough that the tumor escapes containment and the toxic insult of the unchecked tumor on the healthy tissue dominates. At $w=1$ (efficacy-only) the optimum reverts to $N_{\mathrm{doc}}^{*}=3$, as in Section~\ref{sec:single_efficacy}. The interesting observation is that for the entire intermediate range $w\in\{0.3,0.5,0.7\}$ the optimizer agrees on $N_{\mathrm{doc}}^{*}=6$, the value used in clinical practice~\cite{chemprot1, chemprot2}. The Pareto front obtained from the same set of evaluations is shown in Fig.~\ref{fig:pareto}: solutions in the $N_{\mathrm{doc}}\in[3,6]$ range are Pareto-optimal for tumor containment, while $N_{\mathrm{doc}}\in[11,14]$ defines the toxicity-optimal end of the front.

\subsection{Computational efficiency analysis}
Assessing the computational efficiency of Bayesian Optimization is crucial for confirming its practicality in costly, high-dimensional engineering problems.
In scenarios where a single evaluation of the objective function can take several days of computation, conventional hyperparameter tuning strategies such as grid or random search become infeasible. Grid search grows exponentially with the dimensionality of the problem, locking development into a weeks-long cycle of exhaustive trial-and-error. In contrast, Bayesian Optimization (BO) formulates optimization as a sequential decision-making task, relying on a probabilistic surrogate model to explore the search space in a targeted manner. This study shows how the implemented application exploits BO to sharply reduce overall wall-clock time, obtaining near-optimal system configurations in far fewer iterations than those required by unguided search techniques.

To quantitatively assess the efficiency of the proposed method, Table~\ref{tab:efficiency} is reported. In this table, the number of function evaluations required for BO to converge to an optimum (second column) is compared with the number of evaluations required to exhaustively explore the entire search space using a step size of $\Delta(\delta t)=15$ (third column) and, for a finer discretization, a step size of $\Delta(\delta t)=10$ (fourth column).

\begin{table}[h]
    \centering
    \caption{Comparison of the total number of objective function evaluations required for convergence between Bayesian Optimization and exhaustive Grid Search across different therapeutic configurations. The numbers after "+" indicate the number of initial points provider to the algorithm}
    \label{tab:efficiency}
    \begin{tabular}{c c c c}
        \toprule
        Therapy Type & Bayesian Optimization & Grid Search (step$=15$) & Grid Search (step$=10$)\\
        \midrule
        Two therapies & $14+2$ & $\approx14$ & $\approx21$\\
        \midrule
        Three therapies (half - sensitization pos.) & $18+5$ & $\approx11^2=121$ & $\approx17^2=289$ \\
        Three therapies (half - sensitization neg.) & $11+5$ & $\approx11^2=121$ & $\approx17^2=289$ \\
        Three therapies (full) & $13$ & $\approx11^2=121$ & $\approx17^2=289$ \\
        \bottomrule
    \end{tabular}
\end{table}

As indicated by the results, Bayesian Optimization demonstrates a monumental reduction in computational cost, particularly as the complexity of the optimization problem scales. For the two-therapy configuration, BO requires a comparable number of evaluations to the coarser grid search. However, when moving to the three-therapy configurations, the "curse of dimensionality" severely penalizes the exhaustive search approaches. While the grid search combinations explode exponentially from \(14\) to \(121\) iterations (for \(\Delta(\delta t)=15\)) and from \(21\) to \(289\) iterations (for \(\Delta(\delta t)=10\)), the computational footprint of BO remains remarkably flat, requiring at most \(23\) total evaluations. Given that a single function evaluation in this application translates to days of wall-clock processing time, the exponential growth of grid search renders finer discretizations entirely unfeasible in practice, requiring nearly a year of continuous compute. Conversely, by intelligently navigating the search space, the proposed BO framework secures the optimal therapeutic strategy within a highly predictable, manageable timeframe of just a few weeks.

\begin{figure}
    \centering
    \includegraphics[width=1.0\linewidth]{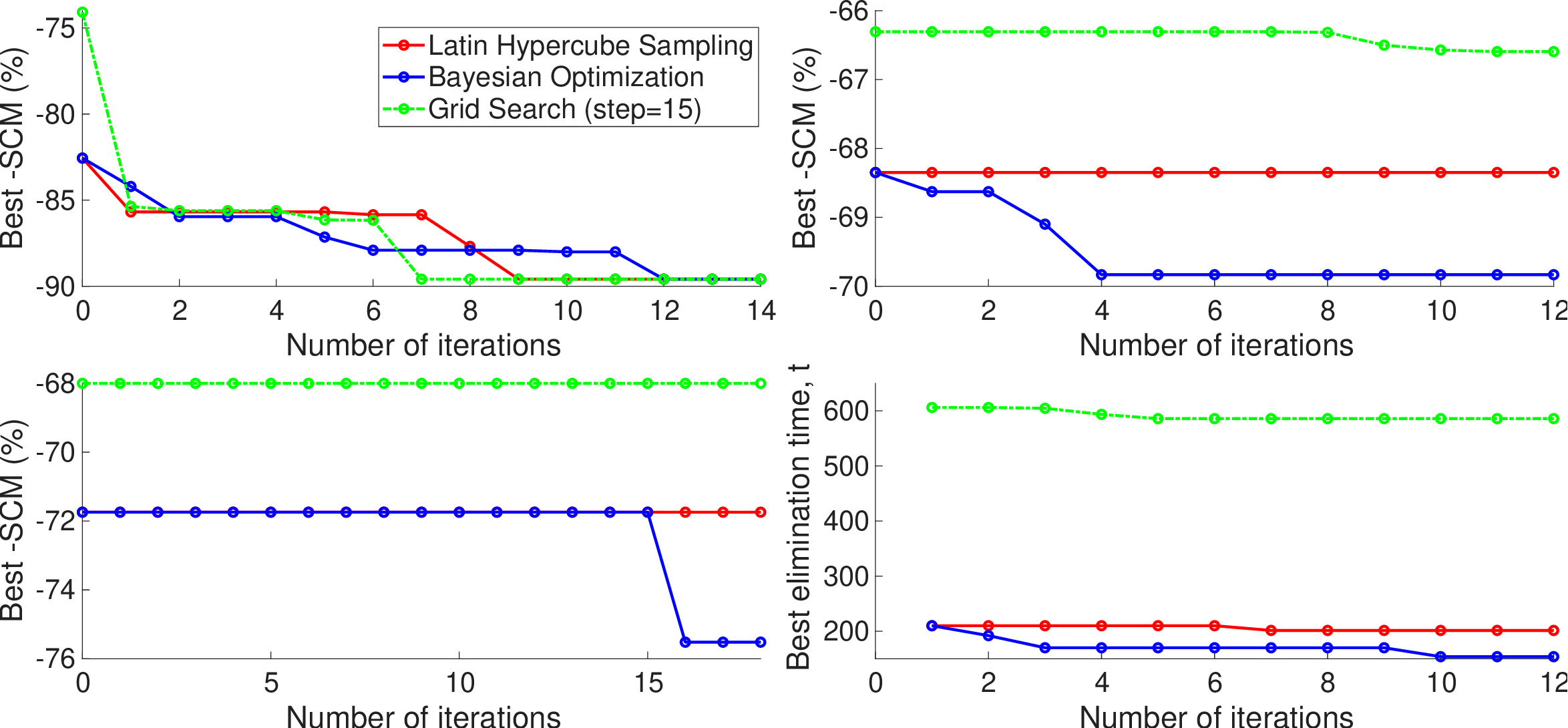}
    \caption{Method efficiency comparison between bayesian optimisation (blue), latin hypercube sampling (red) and, grid search (green) for i. double therapy (top left), ii. triple therapy - sensitization negative (top right), iii. triple therapy - sensitization positive (bottom left), and iv. triple therapy - full regimen.
    For each method, the minimum value obtained is reported under the same iteration budget as that used for BO.}
    \label{fig:methods_compare}
\end{figure}

Complementarily to the above analysis, Fig.~\ref{fig:methods_compare} presents side-by-side comparisons between the bayesian optimization method, the latin hypercube sampling and, the mechanistic grid search (with a  \(\Delta(\delta t)=15\) step). Thus, for the scenarios of double therapy (top left), triple therapy - sensitization negative (top right), triple therapy - sensitization positive (bottom left), and triple therapy with the full regimen, the corresponding optimal objective function values are compared, while maintaining the same number of iterations as those employed by BO.
It is observed that, although Bayesian Optimization exhibits limited efficiency in constrained, one-dimensional search spaces such as those employed in double-therapy configurations, it attains superior efficiency when applied to two-dimensional search spaces.

\begin{comment}
\subsection{Limitations and noise characterization}\label{sec:limitations}

The PDE simulator inherits numerical noise from mesh discretization, time-step adaptivity and nonlinear solver tolerances, and the BO surrogate must absorb this noise to remain well-calibrated. Localized non-monotonicities visible in the surrogates of Figs.~\ref{fig:chapBO_1p_model}, \ref{fig:chap_BO_2p_model_SENS} and~\ref{fig:chap_BO_2p_model_NOSENS} are consistent with this source of variability, and the predictive standard deviation produced by the GP captures it self-consistently. The optimization scripts used to generate the reported surrogate fits are provided in the Supplementary Information (Sec.~\ref{SI-appS}).

We further note that the LCB exploration constant was held fixed at $\kappa=0.1$ across all experiments; the cheaper single-therapy scripts in the Supplementary Information (Sec.~\ref{SI-appS_single}) provide the practical template for repeating this analysis under alternative acquisition parameters.
\end{comment}

\section{Conclusions}\label{sec:Conclusions}

We have presented a Bayesian-optimization framework that couples a multiphase, vascularized PDE tumor model -- implemented as a finite-element solver in COMSOL Multiphysics\textsuperscript{\textregistered} and orchestrated through Python -- with a Gaussian-process surrogate and a Lower Confidence Bound acquisition policy. The framework treats the PDE simulator as a black box and identifies optimal scheduling strategies for combination and single-agent cancer therapies using $\mathcal{O}(10)$ expensive evaluations rather than the $\mathcal{O}(10^2)$ that a comparably resolved grid search would require.

For the docetaxel--bevacizumab schedule, the framework recovers concomitant administration ($\Delta t \approx 0$) as the long-term optimum, in agreement with previous grid-based evaluations~\cite{Lampropoulos2023} and clinical practice. For the triple combination with radiation under reduced intensity, the framework identifies docetaxel-induced radiosensitization as a decisive factor: switching this mechanism off in the model shifts the attained $SCM$ at the global optimum from $75.5\%$ to $69.8\%$ without altering the topology of the optimal region. For the full-intensity triple combination, switching to an elimination-time objective with a penalty for failure reveals a clearly bounded regime ($\Delta t_1\lesssim 20$, $\Delta t_2\gtrsim -20$) in which the tumor is consistently driven to extinction. For the single-agent regimen, the bi-objective formulation recovers the clinically prescribed $N_{\mathrm{doc}}=6$ across a wide range of efficacy--toxicity weightings, providing a quantitative justification of standard practice within the model.

The principal limitation of the present study is the simplified model of healthy-tissue response to cytotoxic agents; a more careful calibration of this response would likely shift the bi-objective optimum toward more fractionated, lower-dose regimens. Other limitations include the use of a single GP kernel family (RBF) and a single acquisition function (LCB), and the deterministic treatment of model parameters; uncertainty quantification through ensemble GPs or Bayesian PDE solvers, integration of patient-specific parameters and dose--volume histogram-based toxicity models, and extension of the framework to true bi-objective acquisition policies (expected hypervolume improvement, ParEGO) are natural directions for future work. The framework itself is agnostic to the choice of PDE model and to the dimensionality of the design space, and we expect it to transfer to other expensive engineering and biological design problems where the cost of a single evaluation precludes brute-force search.

\subsection*{CRediT authorship contribution statement}
\textbf{Ioannis Lampropoulos}: Writing original draft, Visualization, Validation, Software, Methodology.  \textbf{Yorgos M. Psarrelis}: Methodology, Investigation, Writing \& editing original draft. \textbf{Michail Kavousanakis}: Writing original draft, Supervision, Methodology, Conceptualization.

\subsection*{Use of Generative AI statement}
Figure 1, which provides a schematic illustration of the processes captured by the multiphase tumor model, was created with the assistance of Claude. ChatGPT was used solely for language refinement. No AI tools were used to generate, analyze, or interpret the scientific results or conclusions presented in this work.

\subsection*{Declaration of competing interest}
The authors declare that they have no competing financial interests or personal relationships that could have appeared to influence the work reported in this paper. 

\subsection*{Acknowledgements}
M. Kavousanakis would like to thank Professor Y. Kevrekidis (Johns Hopkins University) for insightful discussions on the implementation of the Bayesian Optimization algorithm.

\subsection*{Funding sources}
This research did not receive any specific grant from funding agencies in the public, commercial, or not-for-profit sectors.

%% Loading bibliography style file
%\bibliographystyle{model1-num-names}
\bibliographystyle{cas-model2-names}

% Loading bibliography database

\end{document}